\documentclass[a4paper,11pt]{article}
\pdfoutput=1 

\usepackage{jheppub} 

\usepackage[T1]{fontenc} 

\usepackage{amsmath}
\usepackage{dsfont}
\usepackage{epsfig}
\usepackage{slashed}
\usepackage{bbold}
\usepackage{psfrag}
\usepackage{color}
\PassOptionsToPackage{caption=false}{subfig}
\usepackage{subfig}
\usepackage{multirow}
\usepackage{booktabs}
\usepackage{hyperref}
\usepackage{feynmp}

\newcommand{\fref}[1]{Fig.~\ref{fig:#1}} 
\newcommand{\eref}[1]{Eq.~\eqref{eq:#1}}

\newcommand{\cref}[1]{Chapter~\ref{ch:.#1}}

\newcommand{\nnl}{\nonumber \\}

\newcommand{\beq}{\begin{equation}} 
\newcommand{\eeq}{\end{equation}} 
\newcommand{\ba}{\begin{array}}  
\newcommand{\ea}{\end{array}} 
\newcommand{\bea}{\begin{eqnarray}}  
\newcommand{\eea}{\end{eqnarray} }  
\newcommand{\be}{\begin{eqnarray}}  
\newcommand{\ee}{\end{eqnarray} }  
\newcommand{\bal}{\begin{align}}
\newcommand{\eal}{\end{align}}   
\newcommand{\bi}{\begin{itemize}}  
\newcommand{\ei}{\end{itemize}}  
\newcommand{\ben}{\begin{enumerate}}  
\newcommand{\een}{\end{enumerate}}  
\newcommand{\bc}{\begin{center}}
\newcommand{\ec}{\end{center}} 
\newcommand{\bt}{\begin{table}}
\newcommand{\et}{\end{table}}  
\newcommand{\btb}{\begin{tabular}}
\newcommand{\etb}{\end{tabular}}

\newcommand{\cO}{{\mathcal O}} 
 
\newcommand{\cL}{{\mathcal L}} 
\newcommand{\cl}{{\mathcal L}}

\newcommand{\FdisE}{\ifmmode {\lvert F_{dis}(E_r,E_{int}) \rvert}^2 \else ${\lvert F_{dis}(E_r,E_{int}) \rvert}^2$\fi}
\newcommand{\Fmol}{\ifmmode {\lvert F_{mol}(\mathbf{q,\tilde{\mathbf{q}}}) \rvert}^2 \else ${\lvert F_{mol}(\mathbf{q},\tilde{\mathbf{q}}) \rvert}^2$\fi}

\newcommand{\Mfi}{\ifmmode {\lvert \mathcal{M}_{2-2} \rvert}^2 \else ${\lvert \mathcal{M}_{2-2} \rvert}^2$\fi}

\renewcommand{\subsubsection}[1]{\addtocounter{subsubsection}{1}
\par\nobreak
\medskip
\nobreak
\noindent{\it \thesubsubsection.  #1 }
\par\nobreak\medskip\nobreak}

\def\lpar#1#2#3#4{\rlap{\raise#3\hbox{$\hskip#4#1\left\{\mbox{\phantom{\rule[0mm]{0mm}{#2}}}\right.$}}}
\def\rpar#1#2#3#4{\rlap{\raise#3\hbox{$\hskip#4\left\}#1\mbox{\phantom{\rule[0mm]{0mm}{#2}}}\right.$}}}

\title{\boldmath Phenomenology of a 750 GeV Singlet}

\author[a]{Adam Falkowski}
\affiliation[a]{Laboratoire de Physique Th\'{e}orique, Bat.~210, Universit\'{e} Paris-Sud, 91405 Orsay, France}
\author[b]{Oren Slone}
\affiliation[b]{Raymond and Beverly Sackler School of Physics and Astronomy, Tel-Aviv University, Tel-Aviv 69978, Israel}
\author[b]{Tomer Volansky}

\abstract{
We study the recently reported excess in the diphoton resonance search by ATLAS and CMS.   
We investigate the available parameter space in the combined run-1 and run-2 diphoton data and study its interpretation in terms of  a singlet scalar field which possibly mixes with the Standard Model Higgs boson.  
We show that the mixing angle is already strongly constrained by high-mass Higgs searches in the diboson channel, and by Higgs coupling measurements.
While a broad resonance is slightly favored, we argue that the signal is consistent with a narrow-width singlet which couples to colored and electromagnetically-charged vector-like fermions.   
 Dijet signals are predicted and may be visible in upcoming analyses.   
Allowing for additional decay modes could explain a broader resonance, however, we show that monojet searches disfavor  a large invisible width.   
Finally, we comment on the possible relation of this scenario to the naturalness problem.
}

\begin{document}

\maketitle
\flushbottom

\newpage
\section{Introduction} 
\label{sec:introduction}

The analyses of proton-proton collisions at the center-of-mass energy $\sqrt{s}=13$~TeV at the LHC reveal an excess in the diphoton spectrum near the invariant mass $m_{\gamma \gamma}=750$~GeV~\cite{atlasbump,CMS:2015dxe}.
If confirmed by next year's LHC data or observed in related channels, this will mark a long-overdue discovery of physics beyond the Standard Model (SM). 
The simplest explanation of such a signal is a new boson with $m \simeq 750$ GeV decaying into 2 photons, which, according to the Landau-Yang theorem \cite{Landau:1948kw,Yang:1950rg}, must have spin zero, two, or higher.   
The origin of this resonance, and its possible interactions with the SM are a priori only mildly constrained.

In this paper we discuss theoretical interpretations of the excess, focusing on the spin-0 case.
Indeed, new scalars with  $\cO(100)$~GeV-$\cO(1)$~TeV masses are highly motivated by the fine-tuning problem of the SM.   
Most theories addressing the naturalness problem  predict an extended Higgs sector,
for example the 2nd Higgs doublet in the MSSM, the radial Higgs boson in the twin Higgs model \cite{Chacko:2005pe}, or the radion in the  Randall-Sundrum  model \cite{Randall:1999ee}.
Furthermore, new scalars in this mass range are predicted by Higgs portal dark matter models \cite{Silveira:1985rk} or more general scenarios with hidden sectors \cite{Strassler:2006im}, where they act as mediators between the SM and the hidden sectors.     
The crucial  point is that for new CP-even scalars it is very natural (and in many case inevitable) to mix with the SM Higgs boson. 
Typically, the mixing angle is expected to be $\cO(m_h^2/m_S^2) \approx 0.01$-$0.1$.
 As a consequence, the singlet would acquire the Higgs-like couplings to the SM matter, and the Higgs  boson couplings would be reduced compared to the SM value (see e.g. \cite{Carmi:2012in} for a discussion of the Higgs phenomenology).  
It is important to understand the phenomenological consequences of the mixing and experimental 
constraints on the mixing angle when the new scalar is responsible for the 750 GeV excess.  

To this end, we  first introduce a trivial toy model, where a singlet scalar $S$ couples to a new vector-like quark that carries  SM color and electric charges. 
After integrating out the vector-like quarks at one loop, 
the singlet acquires an effective coupling to photons and gluons.  
Thanks to that coupling, the scalar can be produced at the LHC via gluon-gluon collisions, and it can decay into two photons, much like the SM Higgs boson. 
We show that one can find the parameter space  where the observed 750 GeV excess in ATLAS and CMS can be explained.  
Moreover, the interesting parameter space is not completely excluded by run-1 searches. 
This toy model is a simple existence proof, and it can serve as a building block of more sophisticated constructions.  

Subsequently,  we study the feasibility of the scenario where a singlet $750$~GeV scalar mixes with the Higgs doublet.  
We show that the mixing angle is already strongly constrained by high-mass Higgs searches in the diboson channel, and by Higgs coupling measurements.  
Mixing angles larger than $\sin \alpha \sim 0.1$ are impossible to achieve in this framework. 
This puts a strong constraint on any scenario where the new scalar is somehow involved in electroweak symmetry breaking. 

We also briefly comment on the issue of naturalness. 
While the full  discussion strongly depends on the complete model in which the new scalar is embedded, we point out that the new scalar  may be relevant for this issue. 
Namely, in the parameter space favored by the excess it is  possible that the new scalar and the vector-like quarks take part in cancellation of quadratic divergences of the Higgs boson.

The paper is organized as follows.   In Section~\ref{sec:fit}, we study the ATLAS and CMS diphoton data, identifying the best fit mass, cross-section and width of the proposed scalar.  In Section~\ref{sec:toy} we introduce a minimal toy model which explains the excess via an effective field theory of a singlet that couples to gluons and photons.  The basic features of the model and viable parameter space are identified.   The toy-model is then extended in Section~\ref{sec:DS}, where we allow the singlet to mix with the SM Higgs. 
In Section~\ref{sec:broad} we briefly discuss the implications of a broad resonance followed by the possible constraints and predictions in Section~\ref{sec:exp}.
 We conclude in Section~\ref{sec:outlook} with a discussion of the  possible connection with the naturalness problem.
While this work was in progress, these studies were published \cite{Higaki:2015jag,McDermott:2015sck,Low:2015qep,Bellazzini:2015nxw,Petersson:2015mkr,Molinaro:2015cwg,Ellis:2015oso,Gupta:2015zzs,DiChiara:2015vdm,Franceschini:2015kwy,Pilaftsis:2015ycr,Buttazzo:2015txu,Knapen:2015dap,Nakai:2015ptz,Angelescu:2015uiz,Backovic:2015fnp,Mambrini:2015wyu,Harigaya:2015ezk}. A number of them present ideas that have some  overlap with our study.

\section{A New Resonance?} 
\label{sec:fit}


\begin{figure}[t]
\bc
\includegraphics[width=0.46 \textwidth]{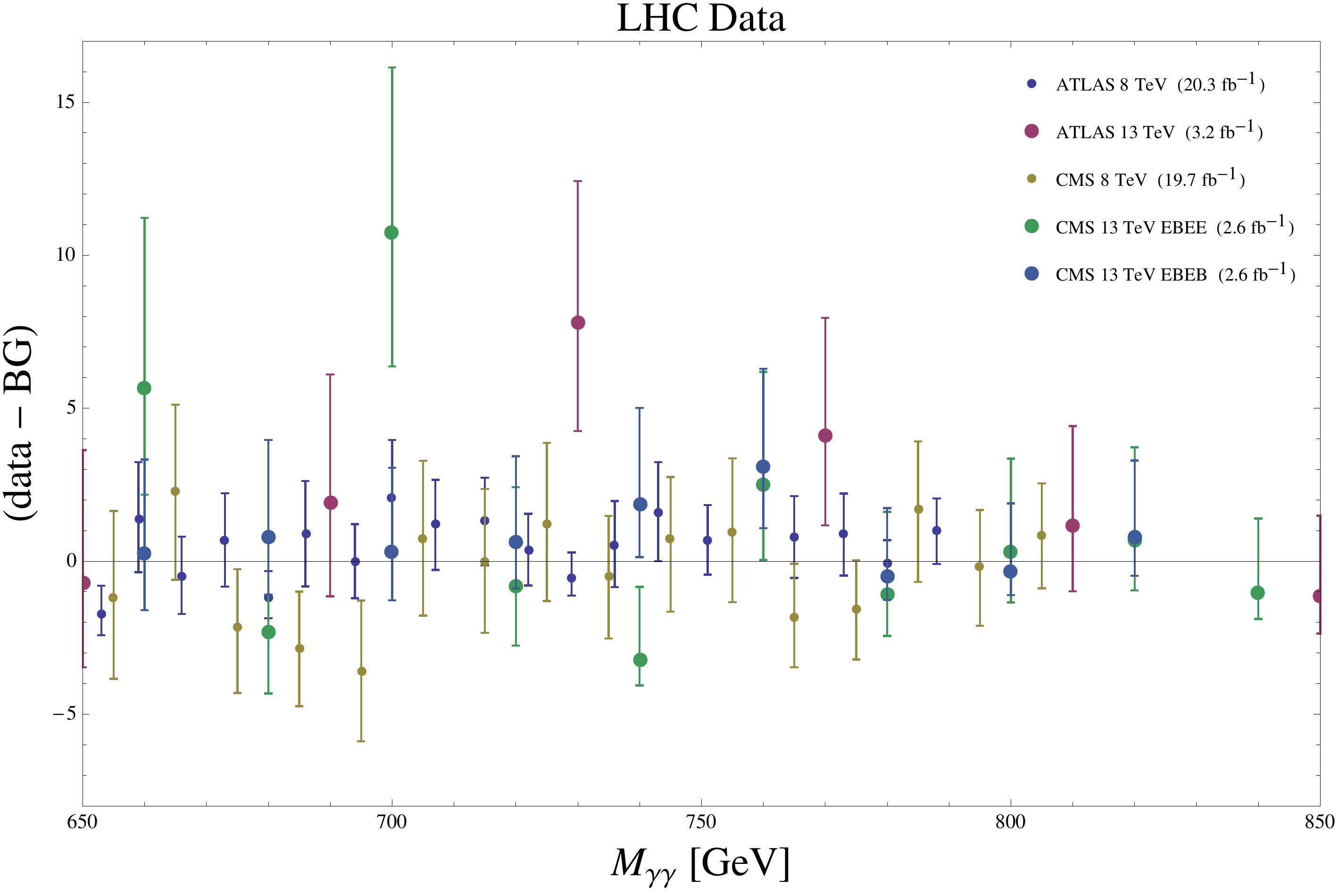}
\quad
\includegraphics[width=0.482 \textwidth]{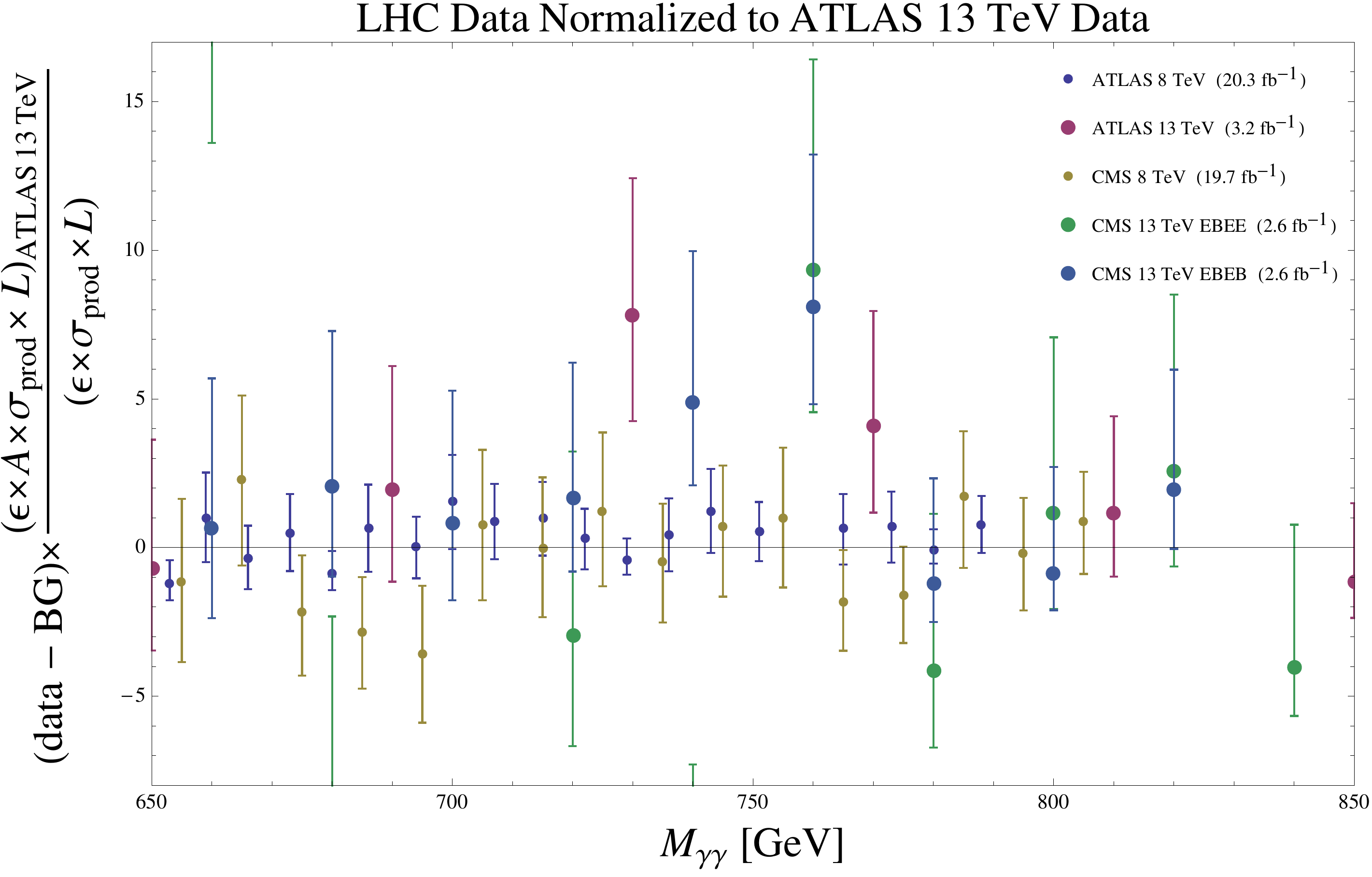} 
\ec 
\caption{
All data sets collected by both the ATLAS and CMS experiments at $\sqrt{s}=\text{8 TeV}$ and $\sqrt{s}=\text{13 TeV}$ runs.  The points correspond to the number of events observed in each bin minus the background fitted functions for each dataset. \textbf{Left:} number of events minus background reported by each analysis. \textbf{Right:} the same data normalized to the ATLAS $13$ TeV cross section, luminosity, acceptance and efficiency. The error bars are normalized as the square root of the data normalization.
\label{fig:data}}
\end{figure} 

The ATLAS experiment presented the diphoton spectrum measured with $3.2~{\rm fb}^{-1}$ collected at $\sqrt{s}=13$~TeV~\cite{atlasbump}. 
In the bins around $750$~GeV, the ATLAS experiment reports the following number of observed events and the estimated SM background prediction: 
\bc 
\begin{tabular} {|c|c|c|c|c|c|c|}
\hline 
Bin[GeV] & 650 & 690 & 730 & 770 & 810 & 850 \\ \hline 
$N_{\rm events}$ &  10 & 10 & 14 & 9 & 5 & 2 \\  \hline 
$N_{\rm background} $ &  11.0 & 8.2 & 6.3  & 5.0 & 3.9 & 3.1  \\  \hline 
\end{tabular}
\ec 
The largest excess is in the bins centered at $730$ and $770$~GeV. 
The local significance of the excess at $750$~GeV is quoted by ATLAS as $3.6~\sigma$. 
There is no evidence for unusual additional activity  (jets, missing energy) in the diphoton events in the excess region, which puts constraints on the production mode of the hypothetical resonance.

CMS uses  $2.6~{\rm fb}^{-1}$ collected at $\sqrt{s}=13$~TeV, and their results are given separately for 2 distinct diphoton categories. In the first category (EBEB) both photons are detected in the barrel,  whereas in the second (EBEE)  one photon is detected in the barrel and the other is found in the end cap.  
The efficiency and acceptance for potential new resonance signals are significantly different in the two categories.
In the bins around 750 GeV they find~\cite{CMS:2015dxe}: 
\vspace{0.5cm}
\bc 
\begin{tabular} {|c|c|c|c|c|c|c|}
\hline 
Bin[GeV] & 700 & 720 & 740 & 760 & 780  & 800
 \\ \hline 
$N_{\rm events}$ (EBEB)&  3 & 3 & 4 & 5 & 1  & 1
 \\  \hline 
$N_{\rm background}$ (EBEB)  &  2.7 & 2.5 & 2.1  & 1.9 & 1.6  & 1.5 
 \\  \hline 
$N_{\rm events}$ (EBEE) &  16 & 4 & 1 & 6 & 2  & 3 
\\  \hline 
$N_{\rm background}$ (EBEE) &  5.2  & 4.6 & 4.0  & 3.5 & 3.1 & 2.8   \\  \hline 
\end{tabular}
\ec 
\vspace{0.5cm}
The EBEB category has a mild excess in the two bins centered at $740$ and $760$~GeV, which coincides with the ATLAS excess. 
The EBEE category (a priori less sensitive)  has a very large excess at $700$~GeV, however without matching signals in the other more sensitive CMS category or in the ATLAS data. The local significance of the excess reported by CMS is $2.6~\sigma$ at around $750$ GeV.

Figure \ref{fig:data} (left) shows the reported data minus background from both experiments at $\sqrt{s}=8$ and $13$ TeV~\cite{Aad:2015mna,atlasbump,Khachatryan:2015qba,CMS:2015dxe}.  Figure \ref{fig:data} (right) presents the same data normalized to the ATLAS $13$ TeV cross section, luminosity, acceptance and efficiency. The normalized CMS $13$ TeV data exhibits better correspondence to the ATLAS $13$ TeV data at around $750$ GeV. 
Both the ATLAS and CMS $8$ TeV normalized data sets  show a mild excess at around $750$ GeV.


In what follows we interpret the reported results in the context of a simple extension of the SM. 
We take a simplified model which includes one additional real scalar, $S$, which has an effective coupling to photons and gluons. In Secs.~\ref{sec:toy}-\ref{sec:DS} we discuss possible models in more detail.  To interpret the above excess we incorporate four distinct data sets. For ATLAS we use the diphoton search at $\sqrt{s}=8$ TeV~\cite{Aad:2015mna} using $20.3$ fb$^{-1}$ of data, and the  $\sqrt{s}=13$ TeV~\cite{atlasbump} search with $3.2$ fb$^{-1}$ discussed above. 
For CMS we take the diphoton searches at $\sqrt{s}=8$~\cite{Khachatryan:2015qba} using $19.7$ fb$^{-1}$ and the $13$~TeV search~\cite{CMS:2015dxe} with $2.6$ fb$^{-1}$.

We work under the assumption that the new particle is  dominantly produced via  gluon fusion.
We mimic a resonant signal using the Breit-Wigner distribution for the scalar mass $m_S \in [700-800]$ ~GeV and the width $\Gamma_S \in [5,100]$~GeV.  
We then perform a Poissonian likelihood analysis in order to find the best fit to the data as a function of three free parameters: (i) the singlet mass, $m_S$, (ii) its width, $\Gamma_S$, and (iii) production times branching ratio rate, $\sigma(pp\to S)\times{\rm Br}(S\to \gamma\gamma)$. 
The production  cross section times branching fraction is scaled by efficiency factors  for each analysis.\footnote{%
For the  $\sqrt{s} = 13$~TeV diphoton analyses, we calculated the efficiency times acceptance for a scalar resonance produced via gluon fusion using Monte Carlo simulated data. 
 For the ATLAS search we find  $\epsilon \times A\approx 0.65$ at $M_{\gamma\gamma}=750$ GeV. 
  For the CMS  search we find  $\epsilon \times A \approx$0.48(0.21) for the EBEB (EBEE) category at $M_{\gamma\gamma}=750$ GeV.
 For the 8~TeV diphoton analyses we use the efficiency times acceptance quoted by  Refs.~\cite{Aad:2015mna,Khachatryan:2015qba}.}
This procedure is applied to the following combined  data sets:
\begin{enumerate}
\item ATLAS $13$ TeV + CMS $13$ (LHC 13 TeV)
\item ATLAS $8$ and $13$ TeV + CMS $8$ and $13$ TeV (LHC 8 and 13 TeV).
\end{enumerate}


\begin{figure}[t]
\bc
\includegraphics[width=0.45 \textwidth]{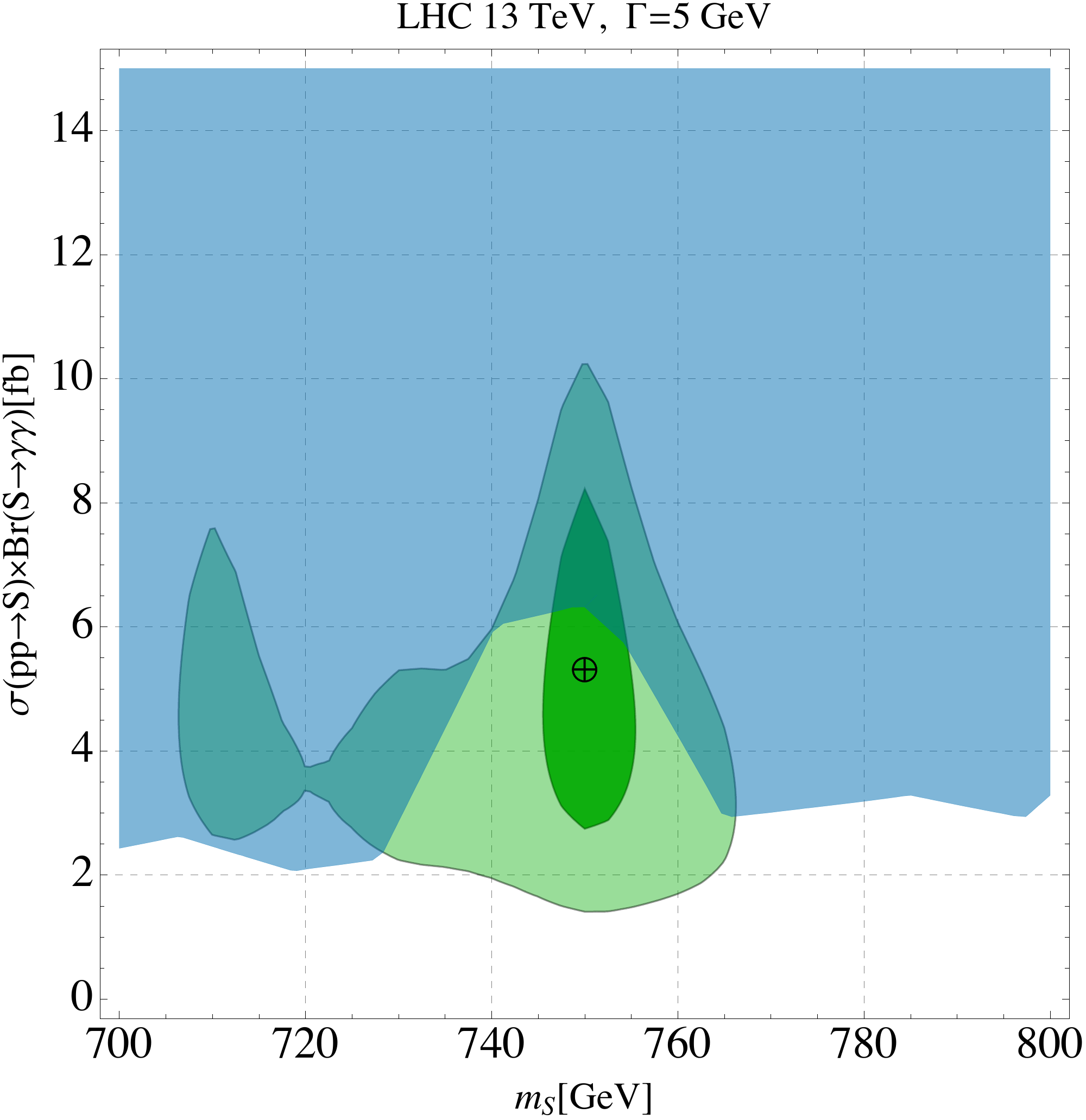}
\quad
\includegraphics[width=0.45 \textwidth]{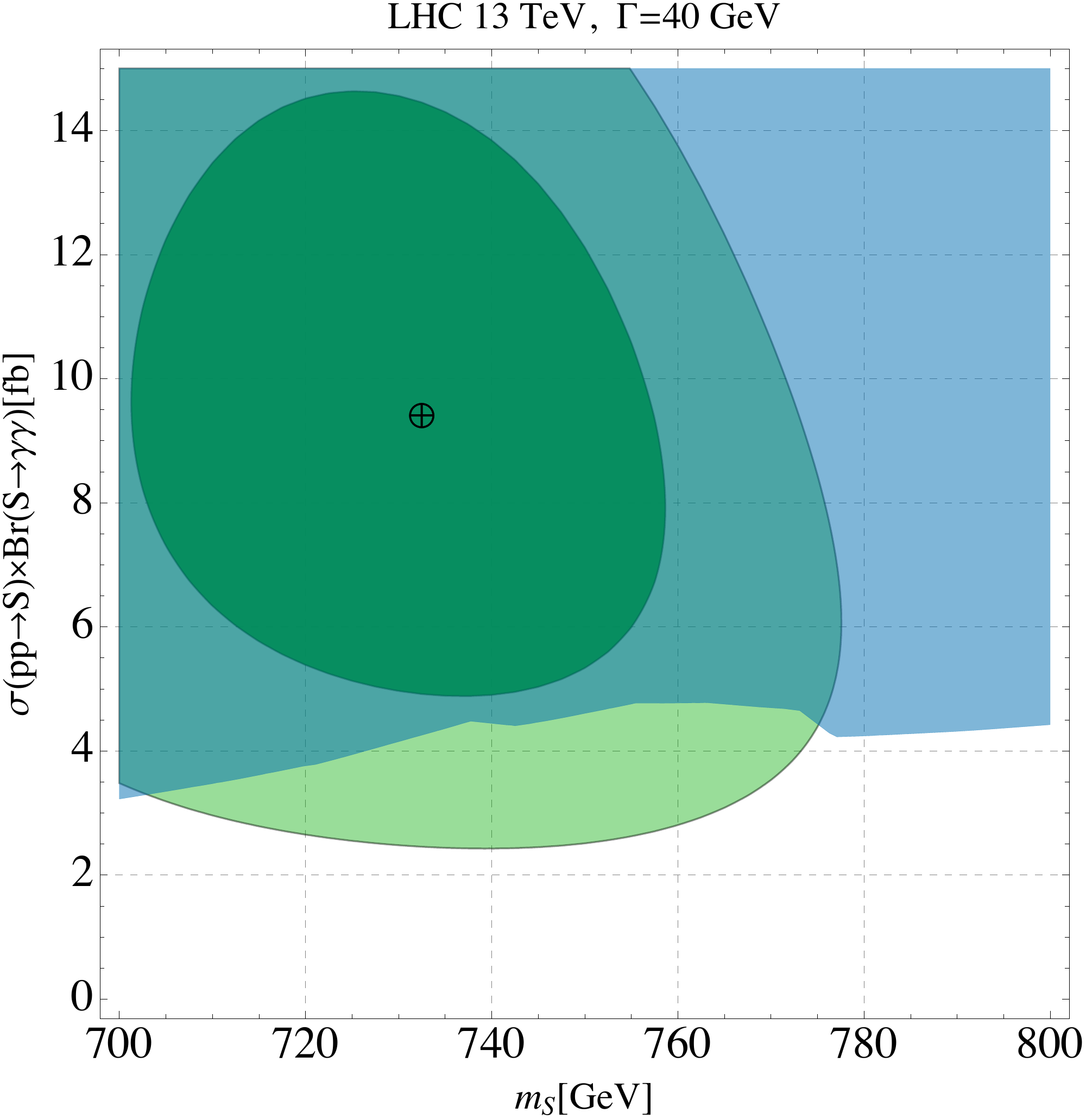} 
\ec 
\caption{
The 68\% CL (darker green) and 95\% CL (lighter green) regions in the plane of mass vs. cross-section of a scalar resonance decaying to 2 photons  favored by  the ATLAS and CMS run-2 data . 
The results are presented assuming a Breit-Wigner shape with $\Gamma_S=5$~GeV {\bf (left)} and 
$\Gamma_S = 40$~GeV {\rm (right)}.  
The blue area is the region excluded by the CMS narrow {\rm (left)} and wide {\rm (right)} scalar resonance search in run-1 \cite{Khachatryan:2015qba}. 
}
\label{fig:bestfit_LHC13}
\end{figure}

\begin{figure}[t]
\bc
\includegraphics[width=0.45 \textwidth]{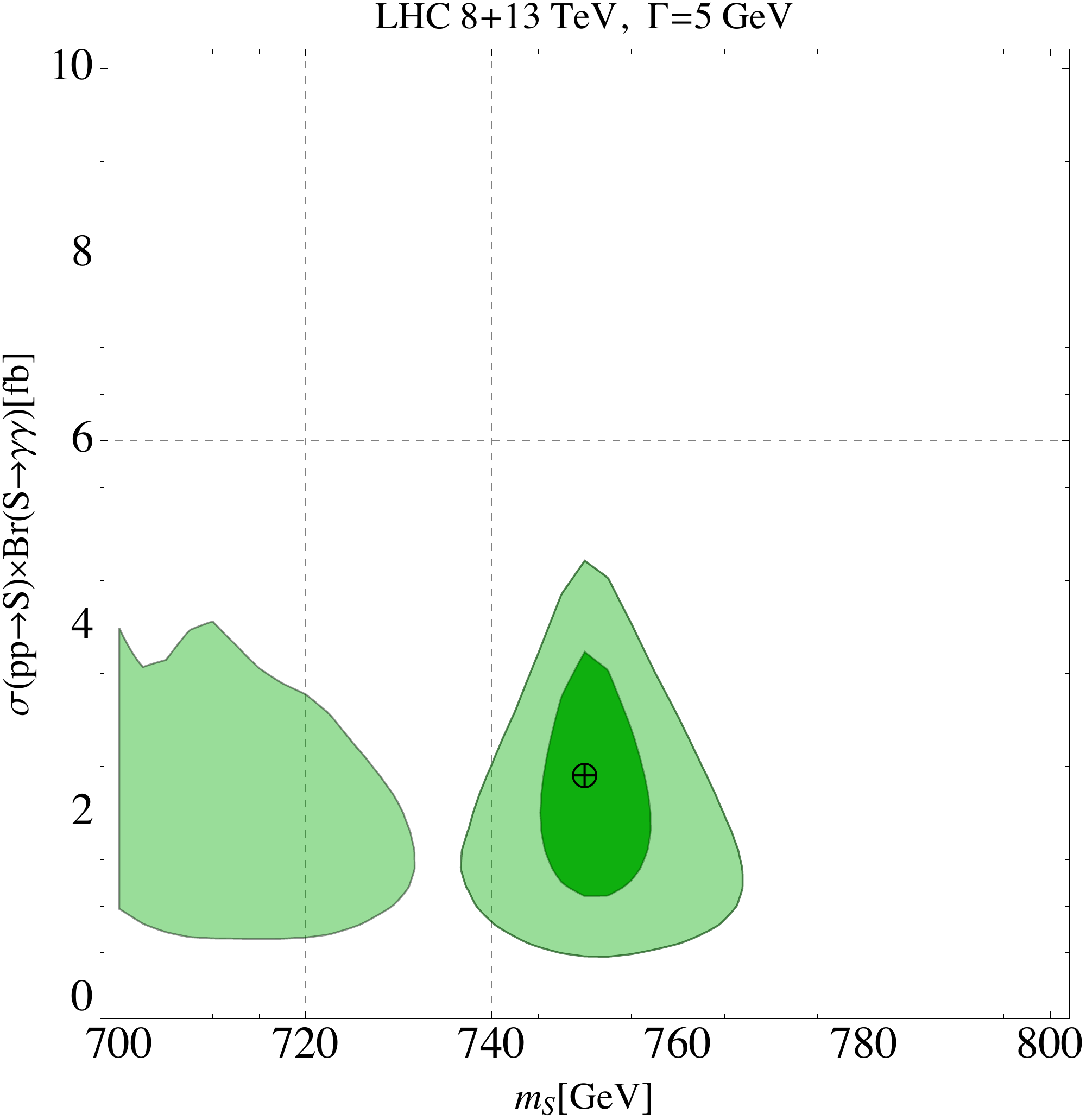}
\quad
\includegraphics[width=0.45 \textwidth]{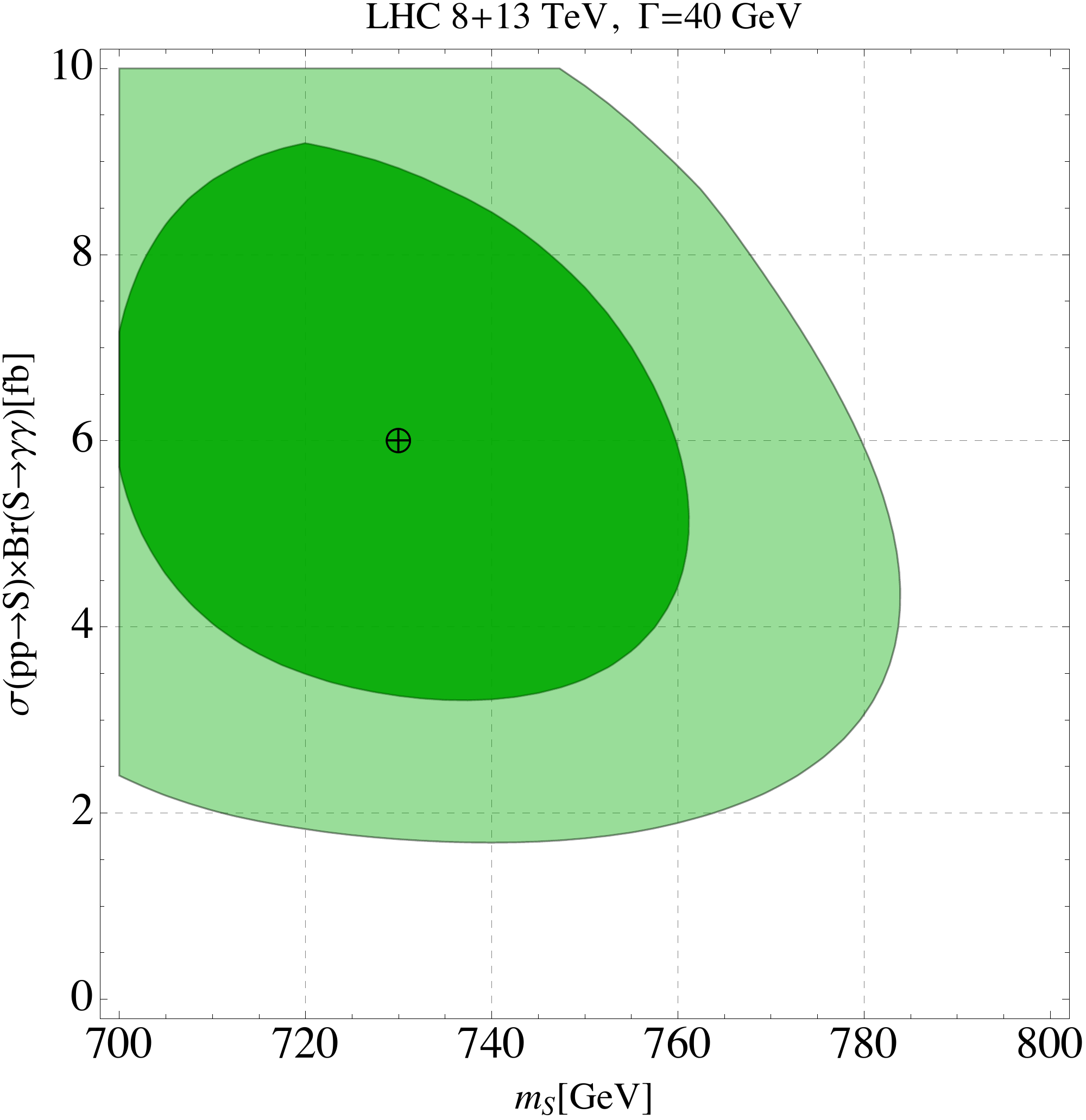} 
\ec 
\caption{
The 68\% CL (darker green) and 95\% CL (lighter green) regions in the plane of mass vs. cross-section of a scalar resonance decaying to 2 photons  favored by  the ATLAS and CMS run-1 and run-2 data . 
The results are presented assuming a Breit-Wigner shape with $\Gamma=5$~GeV (left) and 
$\Gamma = 40$~GeV (right).  
}
\label{fig:bestfit_LHC138}
\end{figure}

The results are presented  in the $m_S$ vs $\sigma(pp\to S)\times{\rm Br}(S\to \gamma\gamma)$ plane for two values of $\Gamma_S=5$~GeV (of the  order of the  the experimental resolution) and $\Gamma_S = 40$~GeV (equal to the bin width in \cite{atlasbump}).  The green colored regions indicate the regions of the parameter space with $68\%$ and $95\%$ CL favored by the data.  Figs.~\ref{fig:bestfit_LHC13} and~\ref{fig:bestfit_LHC138} present results for data sets (1) and (2) respectively.
The values of mass and cross section times BR with the highest likelihood for each data set considered are:

\bc 
\begin{tabular} {|c|c|c|}
\hline 
 & $m_S$ & $\sigma(pp\to S)\times{\rm Br}(S\to \gamma\gamma)$
 \\ \hline 
\textbf{LHC 13 TeV}, $\Gamma$ = $5$ GeV &  $\sim750 \text{ GeV}$ & $\sim 5.3 \text{ fb}$
 \\  \hline 
\textbf{LHC 13 TeV}, $\Gamma$ = $40$ GeV  & $\sim730 \text{ GeV}$ & $\sim 9.4  \text{ fb}$
 \\  \hline 
\textbf{LHC 8 and 13 TeV}, $\Gamma$ = $5$ GeV  & $\sim 750 \text{ GeV}$ & $\sim 2.4 \text{ fb}$
 \\  \hline 
\textbf{LHC 8 and 13 TeV}, $\Gamma$ = $40$ GeV  & $\sim 730 \text{ GeV}$ & $\sim 6.0 \text{ fb}$
 \\ \hline
\end{tabular}
 \label{tab:fit_results}
\ec

If the 13~TeV data only are used, a narrow width resonance with 750~GeV mass and 6~fb cross section provides a very good fit to the data.  There is some tension with the run-1 data, however the best fit point  is not excluded by the  previous diphoton analysis. We do not find any  preference for a large width in the combined ATLAS and CMS data: actually, the best fit points for $\Gamma_S=$5 GeV has $\chi^2$ smaller by one unit than the one for  $\Gamma_S=40$~GeV. 
Once we combine 8 and 13 TeV data, the best fit point moves to a smaller cross section, however the statistical significance of the excess remains high: $\Delta \chi^2 \approx 13$  compared to the SM point for $\Gamma_S = 40$~GeV.
In the combined run-1 and run-2 data a resonance  width larger than the experimental resolution is slightly preferred. 

In \fref{bestfit_LHC138_width} we  also show the best fit region for $m_S = 750$~GeV in the plane of the resonance width vs cross section. 
We see slight preference for a large width: the best fit point occurs for 
$\Gamma_S = 30$ GeV, and $\sigma = 4.8$~fb. 
Finally, in the full 3D scan we find the best fit point for $m_S \approx 730$~GeV, $\sigma \approx 6$~fb,  and $\Gamma_S \approx 40$~GeV. 
This is preferred over the best fit point with $\Gamma_S=5$~GeV by $\Delta \chi^2 \approx 2.5$. 

\begin{figure}[t]
\bc
\includegraphics[width=0.7 \textwidth]{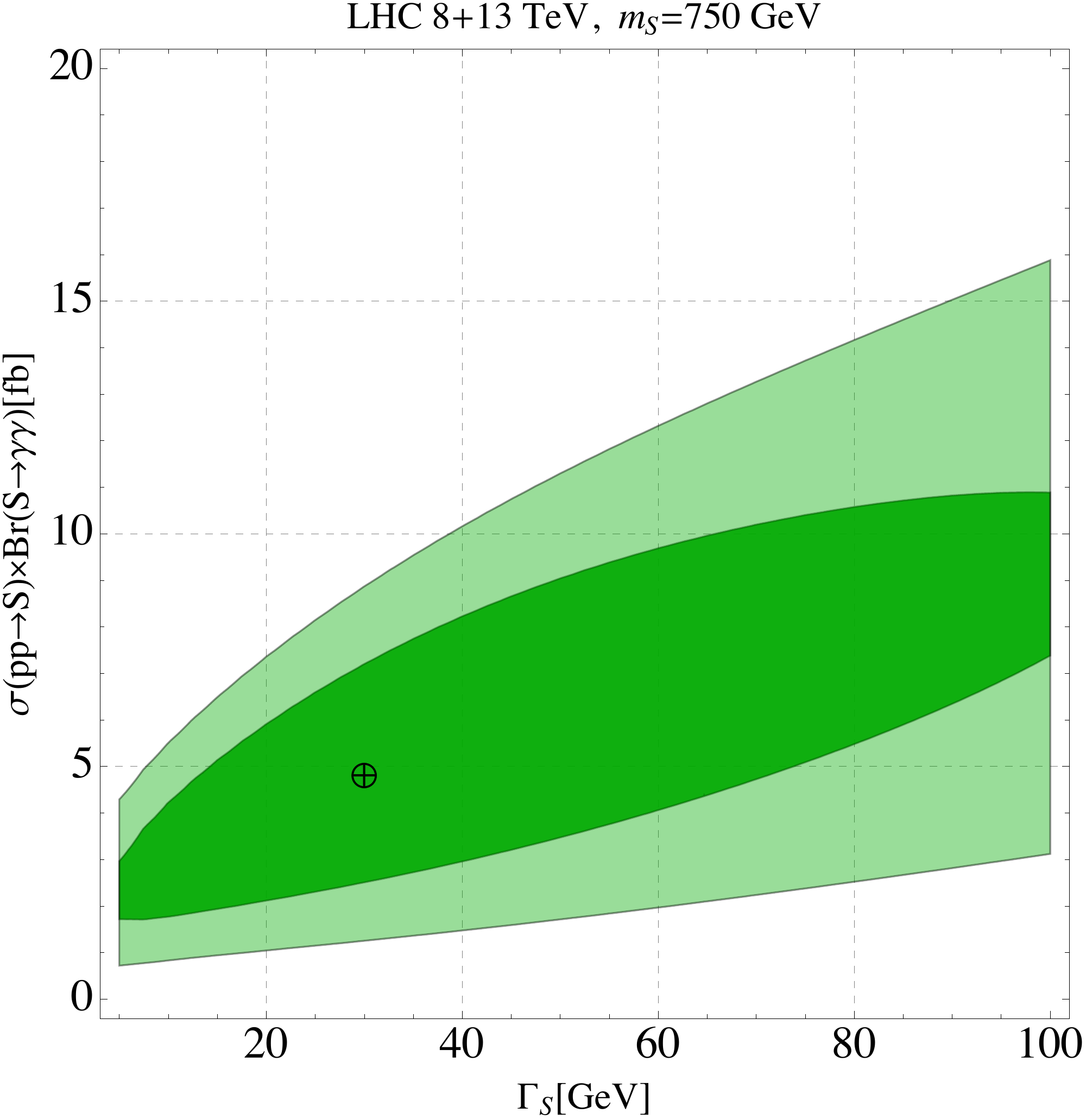}
\ec 
\caption{
The 68\% CL (darker green) and 95\% CL (lighter green) regions in the plane of width vs. cross-section of a 750~GeV scalar resonance decaying to 2 photons  favored by  the ATLAS and CMS run-1 and run-2 data . 
}
\label{fig:bestfit_LHC138_width}
\end{figure}

\section{Toy Model: A Singlet} 
\label{sec:toy}

We begin by studying a minimal  model which addresses the excess discussed above.   
We introduce a real scalar, $S$, coupled to photons and gluons 
\beq
\label{eq:lsgg}
{\cal L}_{S, {\rm eff}} = {e^2 \over 4 v } c_{s \gamma \gamma}  S 
A_{\mu \nu} A_{\mu \nu}
+  {g_s^2 \over 4 v} c_{sgg}  S G_{\mu \nu}^a  G_{\mu \nu}^a, 
\eeq  
where $e$ is the electromagnetic constant, 
$g_s$ is the QCD coupling constant, 
and $v \simeq 246$~GeV is introduced for dimensional reasons. 
In our numerical analyses we use the SM couplings evaluated at $750$~GeV,
$g_s=1.07$, and $e=0.31$. 
These couplings are non-renormalizable, but they 
may arise effectively in a renormalizable model after integrating out vector-like quarks at one loop. 
We assume the singlet has a Yukawa coupling $y_X$ to a vector-like quark $X$ which resides in the fundamental representation of $SU(3)_c$ and has  mass $m_X$ and  electric charge $Q_X$,
\beq
\cL \supset -  y_X  S \bar X  X   \,.
\eeq 
Assuming $m_X \gtrsim m_S$, we integrate out $X$ to generate the following effective couplings to gluons and photons (see e.g. \cite{Carmi:2012yp,Moreau:2012da,Xiao:2014kba,Karabacak:2014nca}):   
\beq
\label{eq:cs}
c_{sgg} =  {y_X v \over 12 \pi^2 m_X} \,, \qquad 
c_{s\gamma \gamma} =   { y_X Q_X^2 v \over 2  \pi^2 m_X}  \,. 
\eeq 
As a consequence, the ratio between the photon and gluon couplings is fixed by the electric charge of $X$, $c_{s\gamma \gamma} = 6 Q_X^2 c_{sgg}$.

\begin{figure}[t]
\bc
\includegraphics[width=0.8 \textwidth]{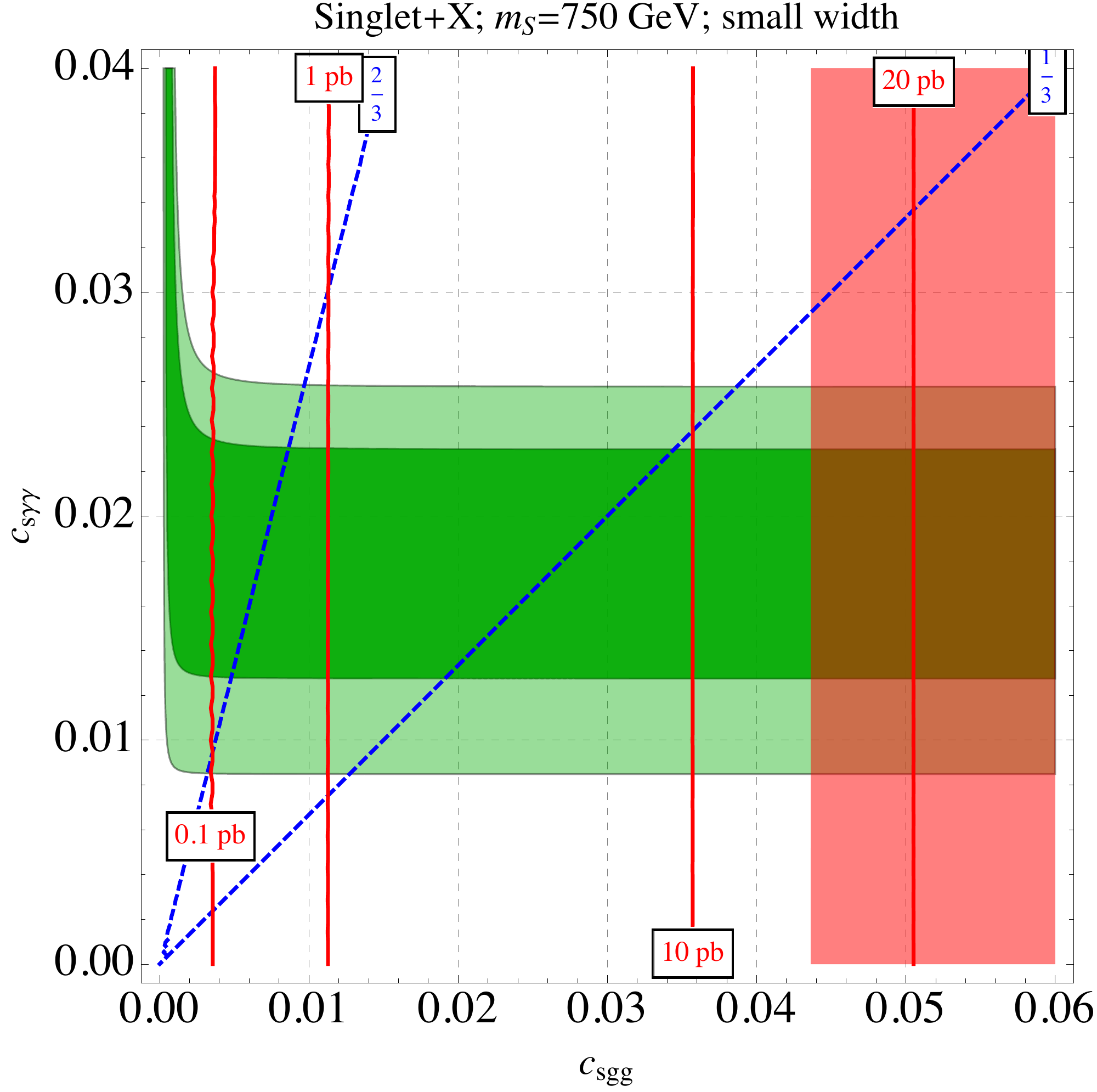}
\ec 
\caption{
The 68\% CL (darker green) and 95\% CL (lighter green) regions of the parameter space favored by  the ATLAS and CMS  diphoton data from run-1 and run-2, assuming a singlet with $m_S = 750$~GeV. 
The blue lines correspond to the couplings generated from integrating out a vector-like  quark with $Q_X = 2/3$ (leftmost) and $Q_X = 1/3$ (rightmost)  interacting via a Yukawa coupling with the singlet.   
The red lines are  contours of constant $\sigma (p p \to g g)$ cross section at $\sqrt{s} = 13$~TeV LHC.
The red-shaded area is excluded by the CMS dijet resonance search at  $\sqrt{s} = 8$~TeV \cite{CMS:2015neg}.
}
\label{fig:singlet}
\end{figure}

The partial  decay widths mediated by these effective couplings are given by,
\beq
\Gamma(S \to \gamma \gamma) =    c_{s \gamma \gamma}^2 {e^4 m_S^3 \over 64 \pi v^2} \, ,  
\qquad 
\Gamma(S \to g g)  =  c_{sgg}^2 {g_s^4 m_S^3 \over 8 \pi v^2}\,.
\eeq 
Assuming that $S$ can decay only to gluons and to photons, 
the branching fraction for the photon decays is found to be, 
\beq
{\rm Br} (S \to \gamma \gamma)  = 
{ e^4 c_{s \gamma \gamma}^2 \over 8 g_s^4 c_{sgg}^2 +  e^4 c_{s \gamma \gamma}^2 }\,. 
\eeq  
Under the same assumption, the total decay width of $S$ is always small for perturbative values of $y_X$ and $m_X \gtrsim 1$~TeV. 
In the narrow width approximation, the tree-level production cross section of the scalar is given by: 
\beq
\sigma(pp \to S) = k { \pi  c_{sgg}^2 g_s^4  m_S^2  \over 64 v^2 E_{\rm LHC}^2 } 
L_{gg}\left ( m_S^2 \over E_{\rm LHC}^2 \right ),
\eeq 
where $L_{gg}$ is the gluon luminosity function, and the $k$-factor accounts for the higher-order QCD corrections. 
The gluon luminosity is obtained using the central value of the NNLO MSTW2008 PDFs  \cite{Martin:2009iq}. 
With this choice, matching to the known NNLO cross sections of a Higgs-like scalar \cite{Heinemeyer:2013tqa}, we estimate $k \approx 3.4$.

With the above, we are now ready to  estimate the range of parameters of the toy model that fit  the ATLAS  and CMS excess, fixing its mass, $m_S = 750$ GeV and assuming a narrow width.
The results are shown in \fref{singlet}.
The best fit regions are obtained assuming $m_S = 750$~GeV contributing to two ATLAS bins at $730$ and $770$~GeV, and to two CMS bins at $740$ and $760$ GeV. 
In the entire displayed region $\Gamma_S \lesssim 1$~GeV, which a posteriori justifies the use of narrow width approximation. 

The experimentally favored region corresponds to the effective coupling to photons in the range $c_{s\gamma \gamma} \in [0.02,0.04]$,  and the couplings to gluon $c_{sgg} \gtrsim 0.01$.  
Clearly,  large Yukawa couplings are needed to arrive at the effective couplings in that ballpark.
For  example, for a vector-like top quark $X$ with $Q_X=2/3$, $m_X = 1$~TeV, and $y_X = 5$ one finds 
$c_{sgg} \simeq 0.01$, $c_{s\gamma \gamma}  \simeq 0.03$.  
Alternatively, one can employ several vector-like quarks with smaller Yukawa couplings. 

In \fref{singlet} we also display the contours of the digluon production cross section at $\sqrt{s} = $~13 TeV, which varies between $\cO(0.1)$~pb and $\cO(10)$~pb  in the interesting parameter space. 
Note that the current run-2 LHC dijet resonance searches \cite{ATLAS:2015nsi,Khachatryan:2015dcf} do not probe the region at $750$~GeV at all. 
 We stress that it is this dijet signal in  the above cross-section range that is the cleanest model-independent verification of this scalar interpretation of the resonance.   
The upcoming ATLAS trigger-level  dijet analysis may therefore shed light on this interpretation.  
The dijet cross-section in run-1 is predicted to be a factor of $\sim 5$ smaller. 
Except for a very large $c_{sgg}$, this is not excluded by the existing run-1 analyses, which set the limit 
$\sigma(pp \to S \to jj) \times A \lesssim 12$~pb  in ATLAS \cite{Aad:2014aqa}
and $\sigma(pp \to S \to gg) \times A \lesssim 1.8$~pb  in CMS \cite{CMS:2015neg}
for $m_S \approx 750$~GeV.
For the CMS search, using parton level simulation we estimate the acceptance $A \approx 0.56$.  
The corresponding dijet cross section at the Tevatron is below a femtobarn, 
and therefore well below the sensitivity of  the CDF search  \cite{Aaltonen:2008dn}.    

Two final remarks  are in order here.   
One is that the results remain unchanged if the singlet scalar is replaced by a pseudo-scalar with the effective couplings \cite{Jaeckel:2012yz}:
\beq
\label{eq:lsggtilde}
\cL \supset {e^2 \over 4 v} \tilde c_{s \gamma \gamma}  S F_{\mu \nu} \tilde F_{\mu \nu}
+  {g_s^2 \over 4 v} \tilde c_{sgg}  S G_{\mu \nu}^a  \tilde G_{\mu \nu}^a\,, 
\eeq
which can be generated by integrating out a vector-like quark with the Yukawa coupling 
$ -  y_X  S \bar X \gamma_5 X$. Then the favored parameter space is still that in \fref{singlet} with the replacement $c_{svv} \to \tilde c_{svv}$. 

The other remark is that, generically, integrating out vector-like quarks at one loop yields effective couplings not only to photons and gluons but also to $ZZ$, $Z\gamma$, and $WW$ \cite{Higaki:2015jag,Buttazzo:2015txu,Franceschini:2015kwy}.
For example, if the vector-like quark has quantum numbers $(3,1)_{Q_X}$  under the SM gauge group then one obtains 
\beq
\label{eq:lsbb}
{\cal L}_{S, {\rm eff}}  \supset {e^2 \over 4 v  c_\theta^2} c_{s \gamma \gamma}  S 
B_{\mu \nu} B_{\mu \nu} = 
 {e^2 \over 4 v } c_{s \gamma \gamma}  S  
\left ( A_{\mu \nu} A_{\mu \nu} 
-  2 t_\theta A_{\mu \nu} Z_{\mu \nu}  
 + t_\theta^2 Z_{\mu \nu}  Z_{\mu \nu}    \right ) , 
\eeq 
where $t_\theta$ are the tangent of the weak mixing angle. 
In this case 
${\rm Br}(S \to Z \gamma)/{\rm Br}(S \to \gamma \gamma) \approx  2  t_\theta^2 \approx 0.6$,
${\rm Br}(S \to Z Z)/{\rm Br}(S \to \gamma \gamma) \approx   t_\theta^4 \approx 0.1$.
On the other hand, if the vector-like quark has zero hypercharge and non-trivial weak $SU(2)$ quantum numbers, then one predicts
 ${\rm Br}(S \to Z \gamma)/{\rm Br}(S \to \gamma \gamma) \approx 2  t_\theta^{-2} \approx 7$, 
 ${\rm Br}(S \to Z Z)/{\rm Br}(S \to \gamma \gamma) \approx   t_\theta^{-4} \approx 13$, 
 and ${\rm Br}(S \to WW)/{\rm Br}(S \to \gamma \gamma) \approx  2 s_\theta^{-4} \approx 40$.
Other patterns may arise when different quantum numbers are assumed, or when the vector-like quarks mix with the SM ones.

\section{The Doublet-Singlet Model} 
\label{sec:DS}

\begin{figure}[t]
\includegraphics[width=0.45 \textwidth]{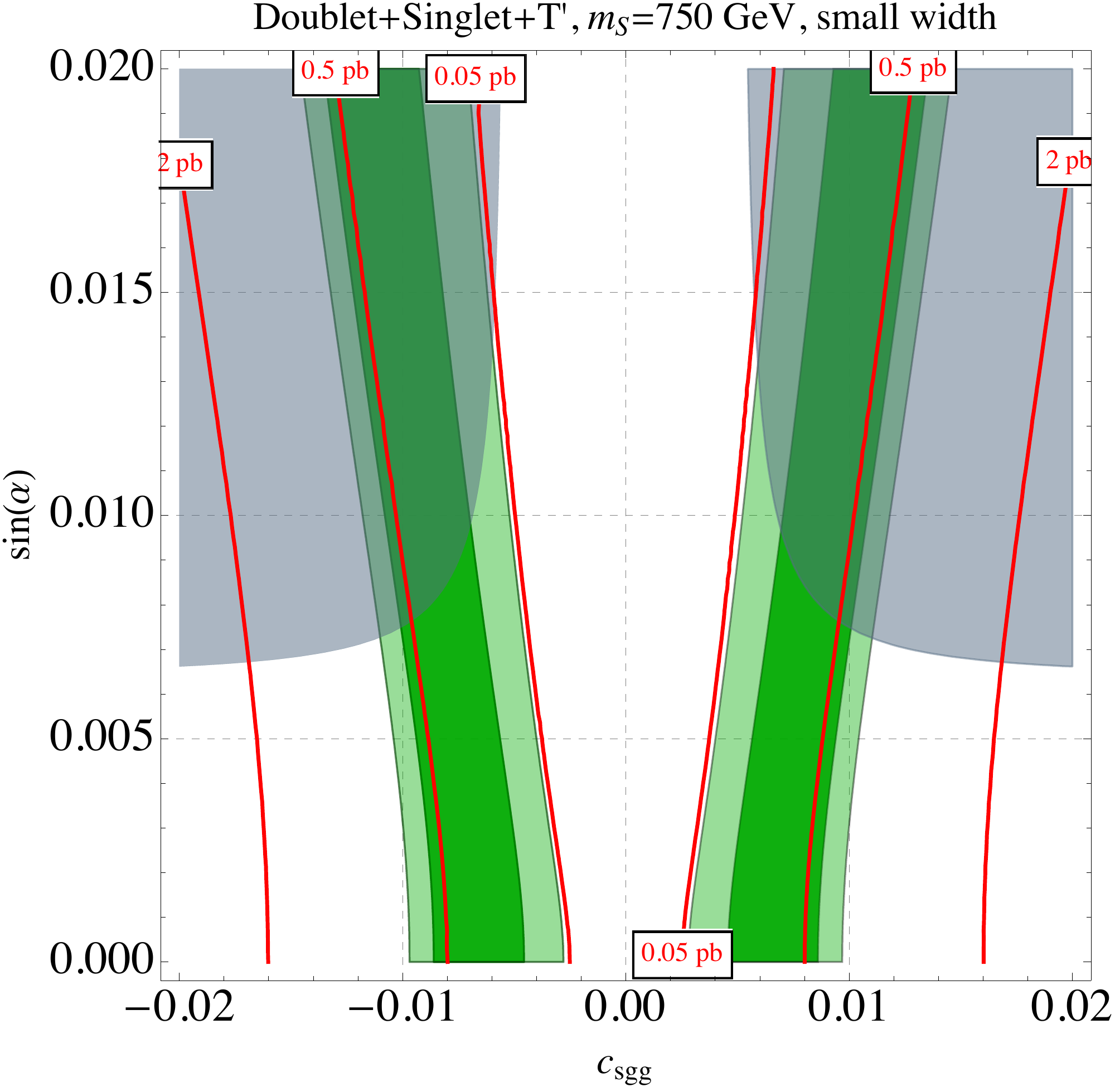}
\quad
\includegraphics[width=0.45 \textwidth]{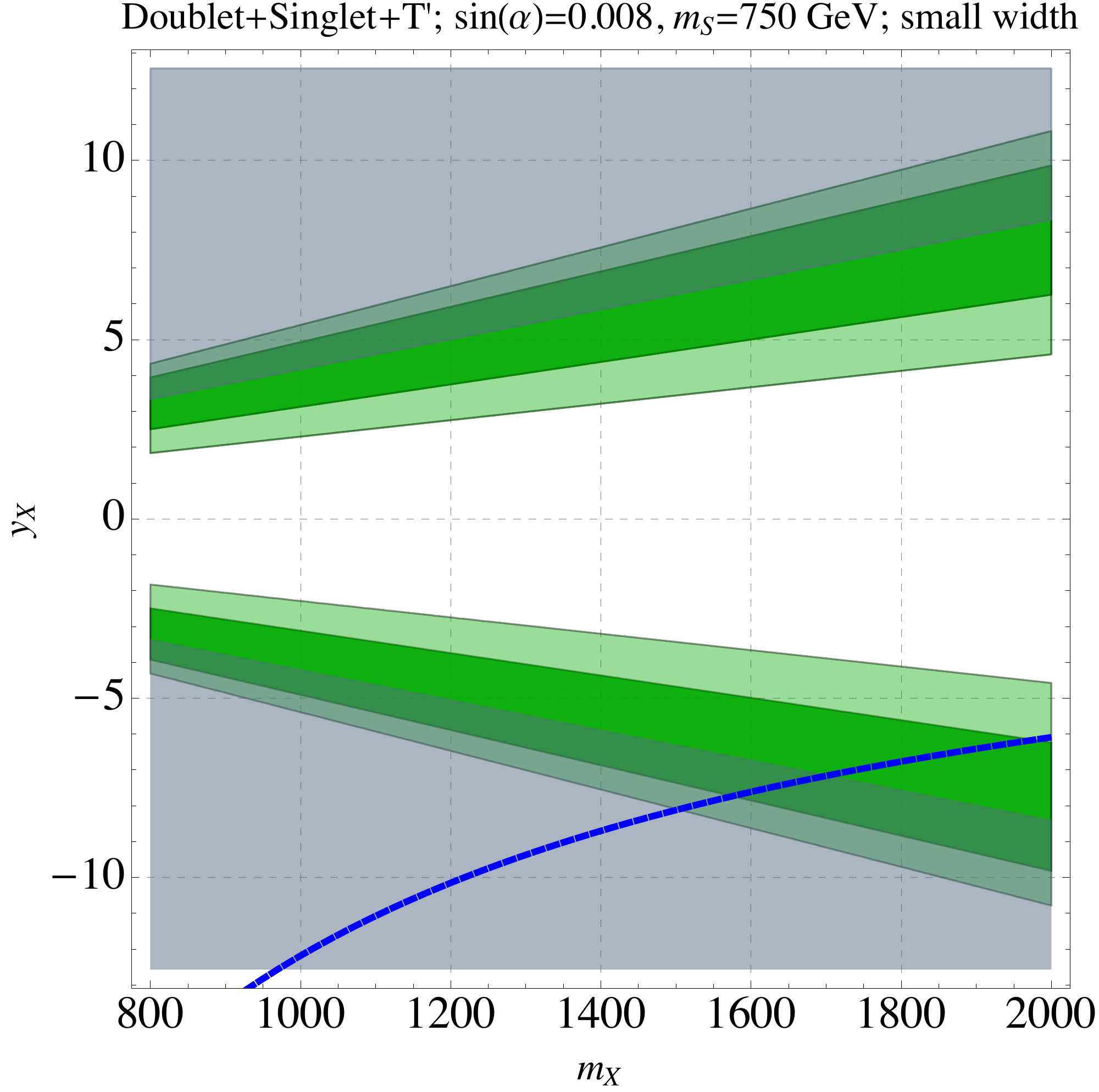}
\quad 
\caption{
{\bf Left:}
The 68\% CL (darker green) and 95\% CL (lighter green) regions of the parameter space of the doublet-singlet model with the singlet coupled to a vector-like quark with charge $Q_X = 2/3$. 
The gray region is the parameter space disfavored at 95\% CL by ATLAS \cite{Aad:2015agg,Aad:2015kna} and CMS \cite{Khachatryan:2015qba} searches for a heavy Higgs boson decaying to WW and ZZ. 
We also show the contours of constant $\sigma(pp \to S) {\rm Br}(S \to gg)$ cross section. 
{\bf Right: } The same  for the mixing angle fixed as $\sin \alpha = 0.008$ and presented in the space of Yukawa coupling $y_X$ and mass $m_X$ of the vector-like quark. 
The dashed blue line marks the parameters for which the vector-like quark cancels the quadratic divergent contribution  to the Higgs mass from the SM top quark. 
}
\label{fig:DS_tprime}
\end{figure}

The singlet model can be extended in a straightforward way.  A particularly motivated scenario is the one in which  the scalar and the SM Higgs boson $h$ mix.  Such a mixing alters the production and decay modes of both the singlet and doublet.  Consequently, precision Higgs measurements and resonant searches in various channels (the most important one being $S\to WW$) place strong constraints on the available parameter space.  

As before, we assume that the singlet couples to new vector-like quarks via the Yukawa coupling 
$\cL \supset - y_X  S \bar X X$.   
In addition, we assume that it couples to the SM  via the Higgs portal, that is through the coupling  $S |H|^2$.   
Then, after electroweak symmetry breaking, the mass matrix of the scalars contains off-diagonal terms.   
To diagonalize the mass matrix one needs to perform the rotation,
\beq
h  \to   h \cos \alpha   + S \sin \alpha \,, \qquad 
 S \to - h \sin \alpha  +  S \cos \alpha  \, ,
 \eeq 
 where $h$ is the physical Higgs mode.
From the Higgs coupling measurements, 
the mixing angle is constrained at 95\% CL to be,  $\sin \alpha \lesssim 0.4$~\cite{lhccomb}, independently of $m_S$.  
Electroweak precision tests impose  slightly stronger constraints in  the relevant mass range:
following the analysis of Ref.~\cite{Falkowski:2015iwa}, 
for a $m_S = 750$~GeV  one finds  $\sin \alpha \leq 0.32$ at 95\% CL.  
In what follows, we  study further constraints on the mixing angle assuming $S$ is responsible for the diphoton excess at $750$~GeV. 
As in the toy model, we find that the decay width of $S$ is always narrow in the relevant parameter space, and therefore the analysis using the narrow width approximation is adequate.

Due to the mixing, the singlet acquires direct couplings to the SM gauge bosons and fermions,
\beq
\label{eq:lh}
\cl \supset 
 {1 \over v} \left ( h \cos \alpha   + S \sin \alpha   \right ) \left [ 
2 m_W^2 W_\mu^+ W_\mu^- + m_Z^2 Z_\mu Z_\mu -  \sum_f m_f  \bar f f \right ] .
\eeq  
This opens the possibility for $S$ to decay to a pair of on-shell $W$ and $Z$ bosons. 
At the same time, the tree-level  Higgs couplings to the SM matter is reduced by $\cos \alpha$, while the one-loop couplings to gluons and  photons  may be altered. 
Furthermore, integrating out the vector-like quark, induces the effective couplings of both $S$ and $h$ to  
gluons and photons: 
\bea 
\label{eq:lsgg2}
\cL & \supset & {e^2 \over 4 v} c_{s \gamma \gamma}  \cos \alpha S  F_{\mu \nu} F_{\mu \nu}
+  {g_s^2 \over 4 v} c_{sgg}  \cos \alpha S G_{\mu \nu}^a  G_{\mu \nu}^a 
\nnl  && 
 - {e^2 \over 4 v} c_{s \gamma \gamma}  \sin \alpha h  F_{\mu \nu} F_{\mu \nu}
-  {g_s^2 \over 4 v} c_{sgg}  \sin \alpha  h G_{\mu \nu}^a  G_{\mu \nu}^a 
\eea 
where $c_{svv}$ are given in Eq.~\eqref{eq:cs}. 
As in the toy model, the couplings in the first line allow $S$ to be produced at the LHC and to decay to photons. 
The couplings in the second line, together with  the modifications in \eref{lh},  affect the Higgs production cross-sections and decay widths.  
We apply the experimental constraints on these couplings from LHC Higgs searches using the likelihood function derived in Ref.~\cite{Falkowski:2015fla}.

In \fref{DS_tprime} we show the results assuming that the vector-like quark $X$ has charge $Q_X=2/3$, that is to say, it is a $T^\prime$ quark with the same color and electromagnetic quantum numbers as the SM top quark. 
We can see that, in this case, the searches for heavy scalars in the diboson decay channel place stringent limits on the mixing angle in the parameters space favored by the diphoton excess,
 $\sin \alpha \lesssim 0.01$. 
 The impact of the Higgs coupling measurements is weaker in the $T^\prime$ case.

\begin{figure}[t]
\begin{center}
\includegraphics[width=0.7 \textwidth]{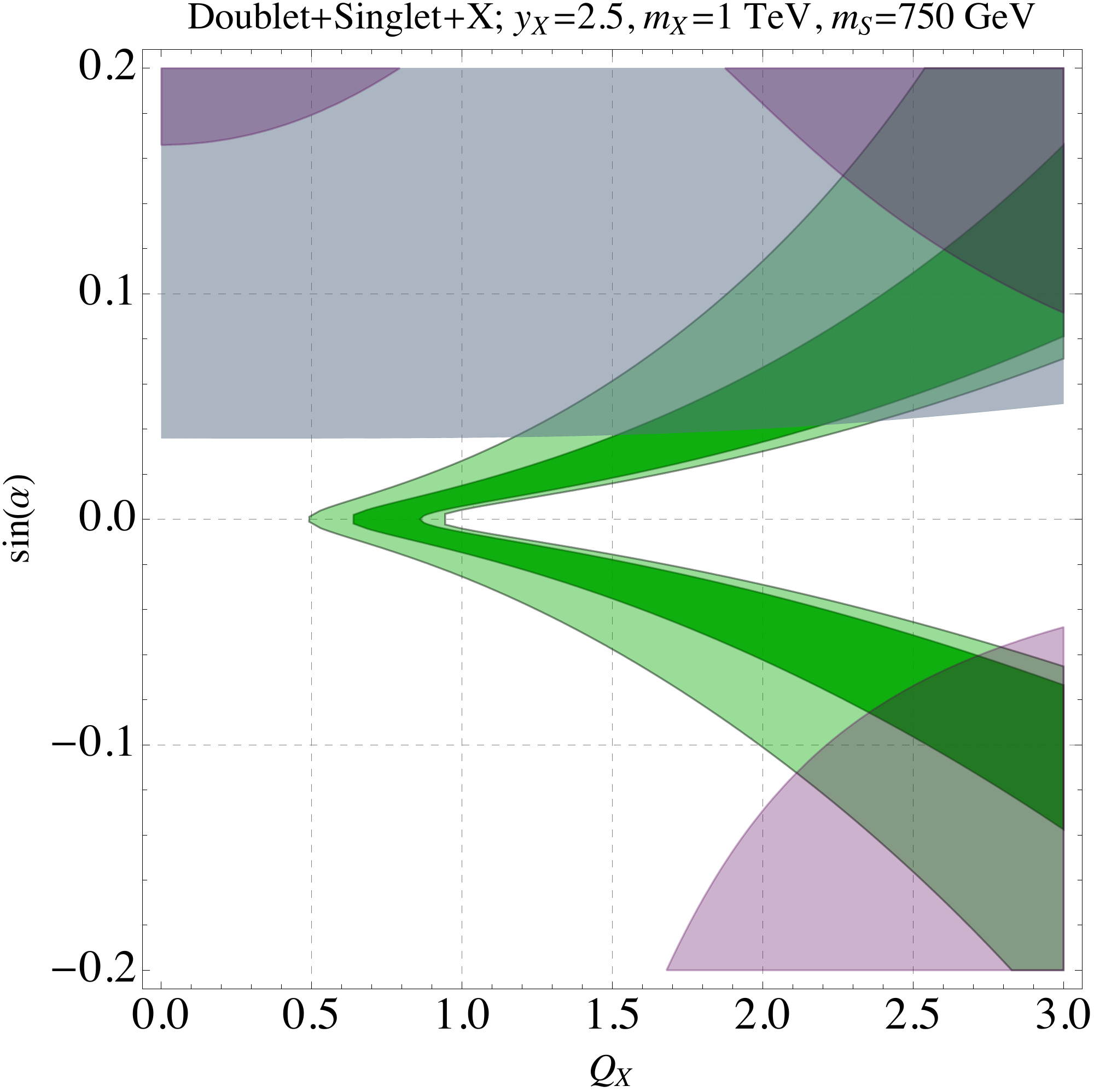}
\end{center} 
\caption{
The 68\% CL (darker green) and 95\% CL (lighter green) regions of the parameter space of the doublet-singlet model in the $\sin\alpha$-$Q_X$ plane, for a fixed Yukawa coupling $y_X=2.5$ and $m_X = 1$~TeV, which corresponds to $c_{sgg} \approx 0.005$.
The gray-shaded region is the parameter space disfavored at 95\% CL by ATLAS \cite{Aad:2015agg,Aad:2015kna} and CMS \cite{Khachatryan:2015qba} searches for a heavy Higgs boson decaying to WW and ZZ. 
The purple-shaded region is disfavored at 95\% CL by Higgs couplings measurements \cite{Falkowski:2015fla}. 
The asymmetry of these two regions between positive and negative $\sin \alpha$ is due to interference between the SM and $X$ contributions to the effective $S$ and $h$ coupling to gluons and photons.  
\label{fig:varQ}}
\end{figure}

\begin{figure}[t]
\includegraphics[width=0.45 \textwidth]{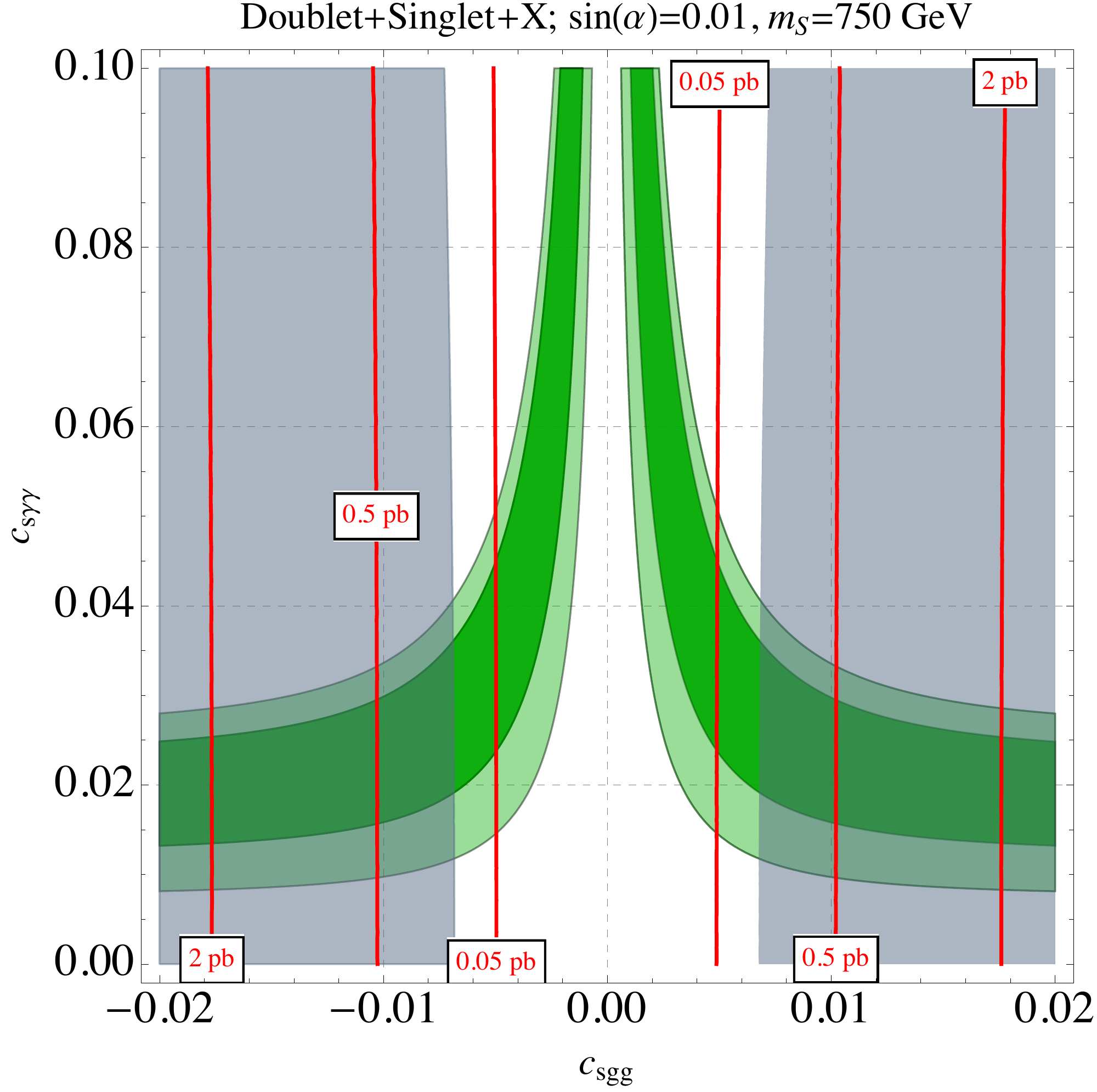}
\quad
\includegraphics[width=0.45 \textwidth]{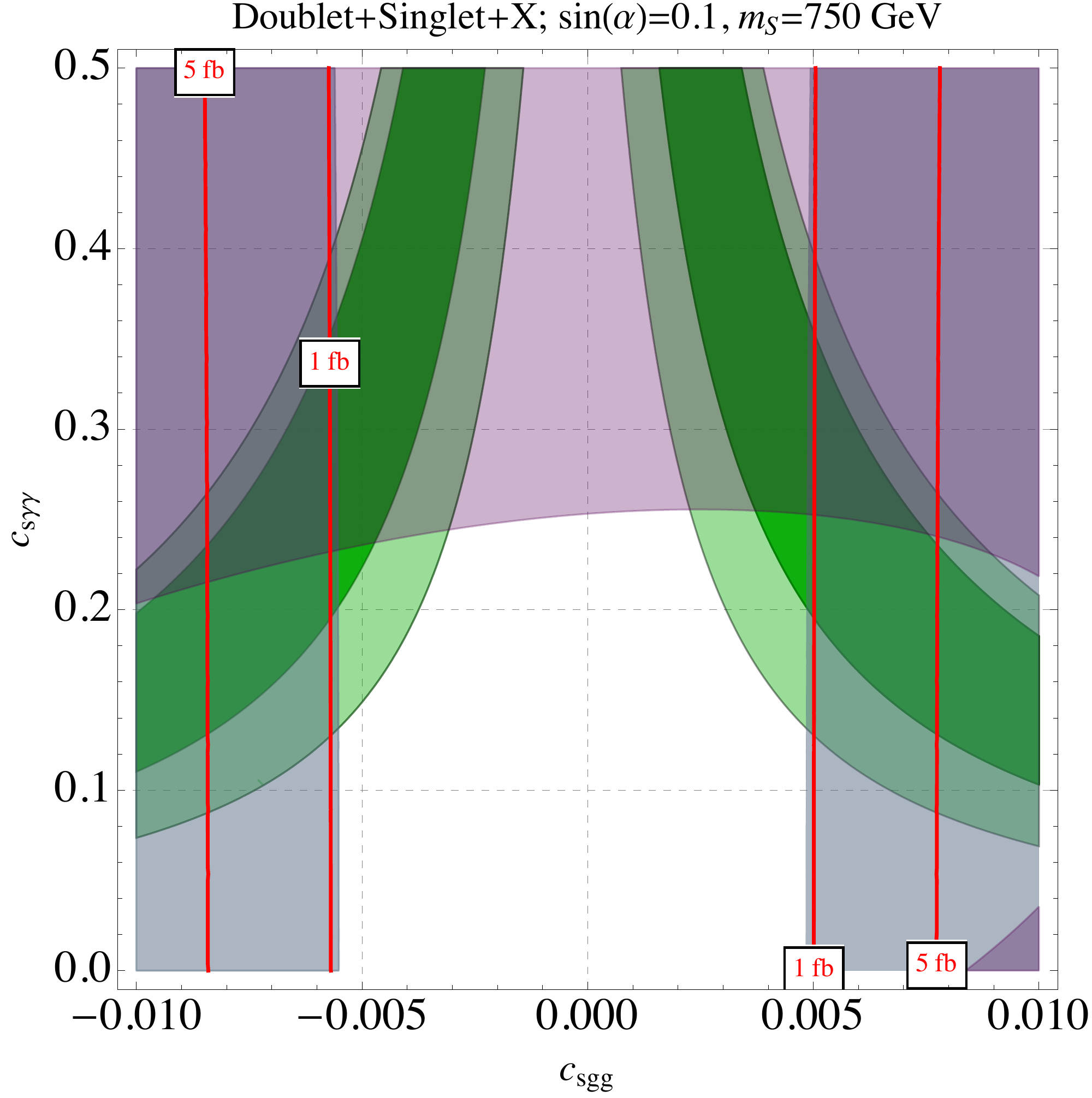} 
\caption{
The 68\% CL (darker green) and 95\% CL (lighter green) regions of the parameter space of the doublet-singlet model in the $c_{sgg}$-$c_{s\gamma\gamma}$ plane, for a fixed value of the mixing angle $\sin \alpha = 0.01$ ({\bf left}) and $\sin\alpha =0.1$ ({\bf right}).
The red lines are  contours of constant $\sigma (p p \to g g)$ cross section at $\sqrt{s} = 13$~TeV LHC.
The gray-shaded region is the parameter space disfavored at 95\% CL by ATLAS \cite{Aad:2015agg,Aad:2015kna} and CMS \cite{Khachatryan:2015qba} searches for a heavy Higgs boson decaying to WW and ZZ. 
The purple-shaded region is disfavored at 95\% CL by Higgs couplings measurements \cite{Falkowski:2015fla}. 
\label{fig:DS_fixedalpha}}
\end{figure}

Changing $Q_X$ or choosing a more complicated pattern of the vector-like quarks, we can change  the relation between $c_{sgg}$ and $c_{s\gamma \gamma}$ compared to the $T^\prime$ case.
This opens up more of the parameter space and allows for larger values of the mixing angle, as shown in Fig.~\ref{fig:varQ} and  Fig.~\ref{fig:DS_fixedalpha}.
We find that the mixing angles as large as $\sin \alpha = 0.1$ can be accommodated in this framework, for sufficiently large Yukawa coupling, $y_X$. 
For larger  $\sin \alpha$ the Higgs couplings measurements  (especially the $h \to \gamma \gamma$ rate) exclude the entire parameter space fitting the 750 GeV excess and still allowed by diboson resonance searches.  
We also note that in the model with $c_{sgg} = 0$, that is when $X$ is a vector-like lepton with no color charge, there is no allowed parameter space at all that fits the 750 GeV excess.  

\section{A Broad Resonance?} 
\label{sec:broad}

As can be seen in Fig.~\ref{fig:bestfit_LHC138_width}, the LHC diphoton data allow for a fairly wide $\cO(10)$-$\cO(100)$~GeV resonance at 750~GeV,  which is significantly larger than the experimental resolution.  
While these hints  are not statistically significant at this point,  it is interesting to contemplate on the implications of such a possibility within the context of the singlet scenario.

We argue that, for the scalar singlet model considered here, and in the absence of additional degrees of freedom that couple to $S$, 
 the width is always narrow in the relevant parameter space, even if the singlet mixes with the Higgs boson.  
Evidence for a wide resonance would therefore hint towards models with a light hidden sector to which the 750~GeV particle  could decay with a sizable branching fraction. 
This sector may or may not be strongly coupled and identifying accompanying  experimental signatures may clarify the situation. 
We distinguish between two distinct possibilities for the origin of the large width:  
\begin{enumerate}
\item $S$ decays invisibly into the hidden sector. 
\item $S$ decays into visible matter, possibly via cascade decays that may or may not involve hidden sector intermediate states. 
\end{enumerate}
In the case (1) the most significant signature is a monojet signal discussed in the next section.   As we will argue then, this possibility is difficult to realize, given the  existing experimental constraints.  Conversely, case (2) predicts additional visible channels which must accompany the diphoton signal.   If a broad resonance is confirmed, such model-dependent visible channels must exist, unless several semi-degenerate states  hide underneath  the observed 750 GeV resonance.

\begin{figure}[t]
\includegraphics[width=0.45 \textwidth]{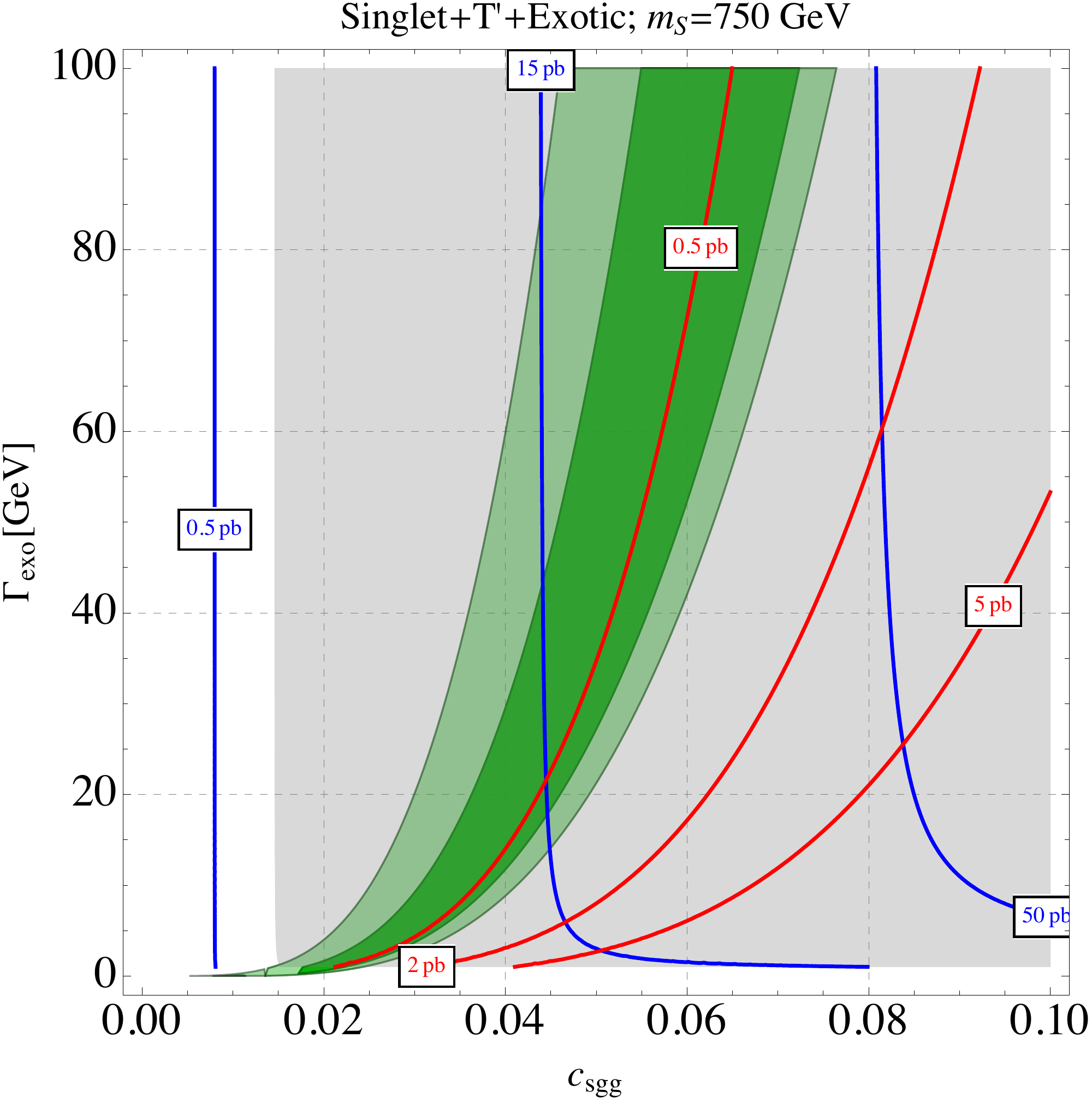}
\quad
\includegraphics[width=0.45 \textwidth]{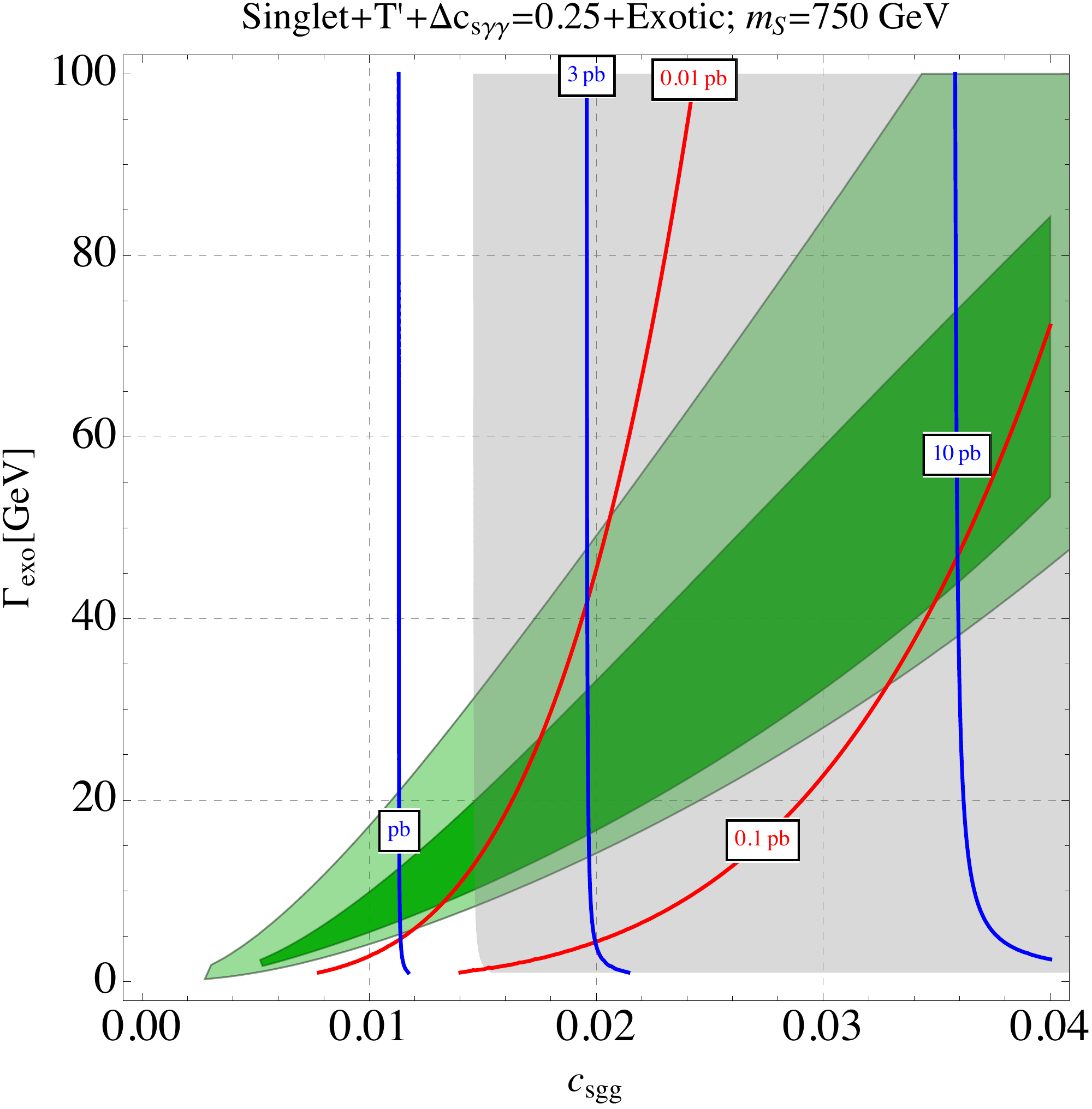} 
\caption{
\textbf{Left:}  Best-fit region for a 750 GeV resonance with a varying width as a function of  $c_{sgg}$, assuming a $T^\prime$ vector-like model.
 Red (blue) lines are the contours of constant digluon (exotic) cross section at the LHC with $\sqrt{s}=$13~TeV. 
 The gray-shaded area is excluded at 95\% CL by ATLAS~\cite{Aad:2015zva} and CMS~\cite{Khachatryan:2014rra}  monojet searches at $\sqrt{s}=8$~TeV, assuming the exotic width corresponds to $S$ decays to invisible particles.  
 \textbf{Right:} The same when, in addition to a $T^\prime$, the model predicts another contribution to the effective coupling of the singlet to photons, $\Delta c_{s \gamma \gamma} = 0.25$. 
\label{fig:inv}}
\end{figure}

Irrespective of what the final state is, a larger width typically implies smaller branching fraction into diphotons. Therefore, the production cross-section must grow as one dials up the total width, in order to fit the LHC excess. Consequently, the rate of exotic processes of the type (1) or (2)  also increases in the interesting parameter space.   
In the left panel of Fig.~\ref{fig:inv}  we demonstrate this effect, by plotting the best-fit region for a 750 GeV resonance with a varying exotic width $\Gamma_{\rm exo}$ as a function of the singlet-gluon-gluon couplings, $c_{egg}$, assuming a $T^\prime$ vector-like model.  
The electric charge of $T'$ sets the gluon-gluon to gamma-gamma ratio, as in Eq.~\eqref{eq:cs}.   
The red and blue contours indicate,  respectively,  the predicted di-gluon and exotic cross sections in the $\sqrt{s}=13$~TeV LHC .
Note that the exotic cross section could also be just dijet, if e.g. the large width is due to large singlet coupling to heavier SM quarks. 
One can see that, in this case, a large width, $\Gamma \gtrsim 20$~GeV, implies an exotic cross section of at least  $O(10)$~pb in the parameter space relevant for the 750 GeV excess.  
In the plot we assume no mixing of the singlet with the Higgs boson, however similar conclusions are reached when mixing is allowed.
The  conclusion can be changed if the ratio between the $\gamma$-$\gamma$ and gluon-gluon effective couplings is larger than that predicted by the singlet model with $T^\prime$. 
In the right panel of Fig.~\ref{fig:inv} we show the situation in a model where, in addition to a $T'$, there are new vector-like leptons providing a very large contribution to the effective coupling of the singlet to photons. As a reference, $\Delta c_{s \gamma \gamma} = 0.25$  corresponds e.g. to 15 vector-like leptons ($\tau'$) with $m_{\tau'} = 750$~GeV and a Yukawa coupling to the singlet, $y_{\tau'} = 3$. 
Nevertheless, even in such an artificially doped scenario, an exotic cross section at least at the picobarn level  is predicted in the favored parameter space with a large width. 
We thus conclude that a search for invisible or exotic signals in run-2 could allow for a  spectacular confirmation of the 750~GeV excess and provide precious new information for model builders.

\section{Experimental Consequences} 
\label{sec:exp}

A statistical confirmation of the excess in the diphoton signal will require the collection of additional data.  Meanwhile, several other channels provide additional constraints and may shed light on the possible nature of the resonance.   
Indeed, $\cO(1)$ diphoton branching fractions are unusual for scalar particles once other channels are kinematically available.  
Even for the 125 GeV Higgs boson, ${\rm BR}(h \to \gamma \gamma)$ is only $0.2\%$, 
while for a would-be 750 GeV SM Higgs boson one has  ${\rm BR}(h \to \gamma \gamma) \sim 10^{-7}$~\cite{Heinemeyer:2013tqa}.
Therefore, it is highly probable that the new particle has other decay channels. 
Below we briefly comment on some existing constraints and possible signals.

\begin{itemize}
\item {\bf Dijets}.
As illustrated in Figs.~\ref{fig:singlet}, \ref{fig:DS_tprime}, \ref{fig:DS_fixedalpha} and \ref{fig:inv}, 
the dijet rate can be significant in the interesting parameter space. 
This is a general  conclusion for scenarios where the 750 GeV resonance is produced in gluon-gluon collisions.
It is therefore crucial to attempt dijet resonance searches in the 700-800~GeV ballpark.  
In the minimal toy model, dijet cross-sections of order 1-10 picobarns in the $\sqrt{s}=13$~TeV LHC can be easily achieved, see Fig.~\ref{fig:singlet}. 
This conclusion also holds for  a large  resonance width, see Fig.~\ref{fig:inv}. 
Once the singlet mixes significantly  with the Higgs boson, constraints from diboson channels preclude a large digluon cross section. 


Current limits from  dijet resonance searches  at  $\sqrt{s} = 8$~TeV   ATLAS~\cite{Aad:2014aqa} and CMS~\cite{CMS:2015neg}  read,  respectively,  $\sigma(pp\to gg) \times A \leq 11$~pb and $\sigma(pp\to gg) \times A \leq 1.8$~pb at 750~GeV. 
Using simulations in Madgraph we estimate the efficiency times acceptance of the CMS search as $A \approx 0.56$.  
The resulting limit does not constrain the available parameter space, except for the large $c_{sgg}$ region in the pure-singlet scenario (see Fig.~\ref{fig:singlet}).  
We expect upcoming trigger-level dijet analysis to improve the limits, possibly allowing to detect the predicted signal in this mass range.

\item{\bf Dibosons}.  
If the scalar mixes with the SM Higgs boson, generically a large signal in the WW and ZZ channels is predicted. 
As discussed earlier,  the ATLAS and CMS  heavy Higgs searches in the WW channel  in run-1 result in   a strong constraint on the mixing angle, $\sin \alpha \lesssim 0.1$. 
Future improvements of the sensitivity in run-2  will  further constrain the parameter space of the model or reveal a signal. 
More generally, in models where scalar decays to $\gamma \gamma$ arise due to loops of heavy particles, decays to $WW$, $ZZ$, and $Z \gamma$ typically occur with rates comparable to the diphoton one. 
For example, if the particle dominating the loop is charged under the SM $U(1)_Y$ and an $SU(2)_L$ singlet, then one predicts ${\rm Br}(S \to Z \gamma) \simeq 0.6 {\rm Br}(S \to \gamma \gamma)$. 
Given the diphoton cross section fitting the LHC data, see \fref{bestfit_LHC138},  
the cross section in the $Z \gamma$ channel is predicted to be $\sim 1$-$2$~fb in the $\sqrt{s} = 13$~TeV LHC,  and  $\sim 0.2$-$0.4$~fb at $\sqrt{s} = 8$~TeV. 
The run-1 ATLAS resonance search in the $Z \gamma$ channel excludes $\sigma (pp \to S) {\rm Br}(S \to Z \gamma) \gtrsim 4$~fb  in the fiducial volume \cite{Aad:2014fha}, which is only an order of magnitude above the predicted signal.   
On the other hand, if the particle dominating the loop is charged under the SM $SU(2)_L$ and a singlet of  $U(1)_Y$, one predicts ${\rm Br}(S \to Z \gamma) \simeq 7 {\rm Br}(S \to \gamma \gamma)$, and then the cross section in the $Z \gamma$ channel is predicted to be of order $1$-$6$~fb at $\sqrt{s} = 8$~TeV.
As shown in \fref{doublet}, this already limits the available parameter space, putting the constraint $\sigma \lesssim 3.5$~fb on the diphoton cross section at  $\sqrt{s} = 13$~TeV for $m_S \approx 750$~GeV.  
Furthermore, for this kind  particle in the loop, the limits from the WW and ZZ channels \cite{Aad:2015agg,Aad:2015kna} are only a factor of 2 weaker.  

\begin{figure}[t]
\bc
\includegraphics[width=0.7 \textwidth]{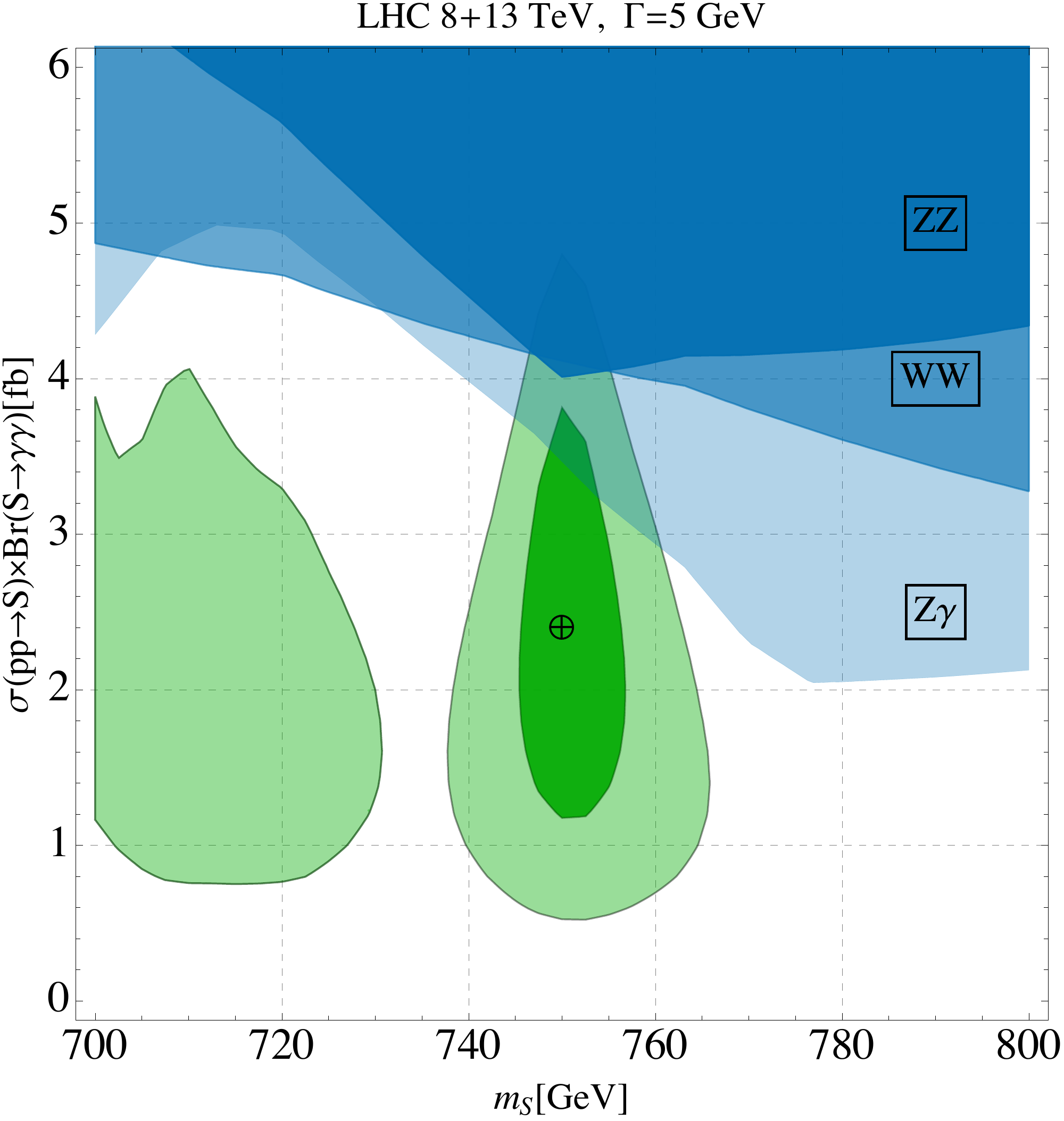}
\ec 
\caption{
The same as \fref{bestfit_LHC138} (left) but with overlaid limits from ATLAS $Z\gamma$ \cite{Aad:2014fha}, $WW$ \cite{Aad:2015agg}, and $ZZ$ \cite{Aad:2015kna} resonance searches (different shades of blue)  at $\sqrt{s}=8$~TeV, assuming the particle generating the effective coupling of the singlet to photons and gluons has zero hypercharge.
For the $Z\gamma$ limits, we took the acceptance in the fiducial volume $A \approx 0.7$ for $m_S \approx 750$~GeV, estimated using Madgraph simulations of the $p p \to S \to (Z \to  \ell^+  \ell^-) \gamma$ process. 
}
\label{fig:doublet}
\end{figure}

\item{\bf New colored states}.
New particles (fermions or bosons) with a color and electric charge and a large coupling to the 750 GeV singlet scalar are needed to generate an effective coupling of the scalar to gluons and photons.
These new colored states should not be too heavy, otherwise their couplings to the scalar must enter the non-perturbative regime in order to explain the diphoton excess. 
As can be seen in \fref{DS_fixedalpha}, in a model where a new vector-like top quark is the only new colored state, its Yukawa couplings to the scalar must already be $y_{T'} \gtrsim 3$ for $m_{T'} \approx 1$~TeV.
Therefore, if this interpretation of the diphoton excess is correct, one expects new colored states just around the corner.  
These new states may decay to the SM (if they mix with the SM matter) or they can be stable on the collider scale in which case the search for heavy stable states and R-hadrons become important.

\item{\bf Monojets}.
Following the discussion in the previous section, it is a priori possible to explain a broad resonance through decays of the scalar to invisible particles,  $S \to \chi \chi$.
In Fig.~\ref{fig:inv}, the blue contours should then be interpreted as an invisible cross section, $\sigma(p p \to \chi \chi)$ at $\sqrt{s}=13$~TeV. 
That is not directly observable in a collider, however the associated monojet signature  can be observed, when the invisible particles recoil against an energetic jet  emitted in the process.  
Limits on this scenario may be derived from run-1 monojet searches in ATLAS~\cite{Aad:2015zva} and CMS~\cite{Khachatryan:2014rra} which put constraints on the monojet rates with $p_{T,\rm miss}$ ranging from $150$ to $700$~GeV.   
As an example, the monojet cross section, $\sigma (p p \to \chi \chi g)$ with $p_{T,\rm miss}>500$~GeV at $\sqrt{s}=8$~TeV, calculated at the leading order using Madgraph, is a factor $3 \times 10^{-3}$ smaller than the invisible cross section.  
For this missing energy bin, the ATLAS (CMS) quote the constraint $\sigma \lesssim 7(6)$~fb.  
On the other hand,  in the best-fit region of the $T^\prime$ model where the resonance is broad  one predicts  the monojet rate well above  $10$~fb, which is excluded (c.f. the gray-shaded region in the left panel of Fig.~\ref{fig:inv}.)  
We thus conclude that it is problematic to address the broadness of the resonance by assuming a large invisible width. 
Instead, if the hints of  $\cO(10-100)$ GeV width are confirmed, 
new visible (though possibly exotic) signals would be expected to accompany the diphoton signal. 
This conclusion can be circumvented if a new large contribution to the effective coupling to photons is present, as in the right panel of Fig.~\ref{fig:inv}. 
However, huge contributions are required to this end, possibly pointing to UV completions by a large $N_c$ strongly interacting theory \cite{Franceschini:2015kwy}.

\item{\bf Hidden valley and lepton jets}.  
Given the strong constraint from monojets on the invisible decay width, it is possible that  $S$ cascades to additional hidden-sector particles which then, at least in part,  decay visibly.  This scenario is referred to as  the {\em hidden valley} models~\cite{Strassler:2006im} and their precise signature is strongly model-dependent.   
Here we mention one interesting possibility of decays to lepton jets \cite{ArkaniHamed:2008qp}, which can be e.g. realized in a similar model as in Ref.~\cite{Falkowski:2010gv}.   
Such decays may occur promptly or induce displaced vertices, depending on the corresponding parameter space.    Limits on such a scenario have been studied mostly  for a light scalar and cannot be straightforwardly interpreted for this scenario (see e.g.~\cite{Aad:2014yea, Aad:2013yqp, Aad:2012kw, ATLAS:2014kla,Khachatryan:2015wka,Aad:2015sms,Aad:2012qua}).   
Additional searches are thus required and may place interesting (even if model-dependent) limits. 
 
\end{itemize}

\section{Outlook: Naturalness Around the Corner?} 
\label{sec:outlook}

In this paper we argued that the SM supplemented by a 750 GeV singlet scalar and an $\cO(1)$~TeV vector-like quark with a  sizable  Yukawa coupling to the singlet can very well explain the observed 750 GeV diphoton excess. 
This explanation continues to be valid if the singlet has a small mixing $\sin \alpha \lesssim 0.1$ with the SM Higgs boson. 
While a broad-width resonance  fits the data better, a narrow resonance is still consistent with the present data.    Given the strong available  monojet constraints, if a broad resonance is confirmed it is challenging to explain 
with a sizable invisible width.  Consequently, accompanying model-dependent visible channels are expected, unless underneath the resonance lie a set of degenerate states.   In addition, we find that upcoming dijet searches will allow for an exciting model-independent confirmation of the singlet interpretation to the data.

From the purely phenomenological view point, 
it is  possible that this simple extension constitutes the complete theory of fundamental interactions at the TeV scale. 
Nevertheless,  a reasonable expectation is that these degrees of freedom are just a tip of an iceberg, revealing the hints of a more sophisticated structure of physics beyond the SM. 
An exciting possibility is that this larger structure is responsible for solving the hierarchy problem of the SM. 

The diphoton excess may hint to such a possibility.  
Indeed, if the resonance couples to photons and gluons via one-loop diagrams, it must couple to colored and electrically charged particles.   Such states may then be involved in canceling the top-induced quadratically divergent contribution to the Higgs mass.     In the case of a mixing between the singlet and Higgs doublet, such colored states couple to the Higgs and may result in the desired cancelation.   It is therefore interesting to ask whether such a cancelation occurs in the allowed  parameter space.    The blue line in Fig.~\ref{fig:DS_tprime} shows the parameters in the singlet-fermion Yukawa coupling vs.~fermion mass plane where such a cancelation occurs.  Interestingly, this possibility is not excluded and implies that the observed resonance may be a first hint of naturalness.   Closing in on this region and identifying the accompanying channels may therefore enable us to progress towards a solution to the naturalness problem.

\section*{Acknowledgement}

We thank Patrick Meade, Yossi Nir, Erez Etzion, Liron Barak and Kohsaku Tobioka for fruitful discussions which contributed to this study.
A.F. is supported by the ERC Advanced Grant Higgs@LHC. 
O.S. and T.V. are supported in part by a grant from the Israel Science Foundation. T.V. is further supported by the German-Israeli Foundation (grant No.~I-1283-303.7/2014) and by the I-CORE Program of the Planning Budgeting Committee and the Israel Science Foundation (grant No.~1937/12).

\bibliographystyle{JHEP} 
\bibliography{singlet750}


\end{document}